\documentclass[a4paper,12pt]{article}
\usepackage{amssymb,amsmath,amsfonts,placeins,pbox,multirow,mathtools,bbold}
\usepackage{graphicx,rotate,color,slashed,cite,epstopdf,verbatim,url,multirow,booktabs}
\usepackage{epstopdf}
\usepackage{subfig}
\usepackage{graphicx}
\usepackage{array}
\renewcommand{\arraystretch}{1.5} 

\usepackage[colorlinks=true,
            linkcolor=magenta,
            urlcolor=blue,
            citecolor=blue]{hyperref}
\usepackage[utf8x]{inputenc}

\numberwithin{equation}{section}

\newcommand{\tr}{\mathrm{tr}}

\newcommand{\complexcos}{\rm SU(4)_l\times SU(4)_r/SU(4)_d}
\newcommand*{\thead}[1]{\multicolumn{1}{c}{\bfseries #1}}

\def\ssbh#1//#2//{\ensuremath{\xrightarrow [\substack{#2}]
    {\parbox{3cm}{\hfil $\scriptstyle \langle #1 \rangle$ \hfil}}}}

\let\CapTion=\caption
\def\caption#1{\CapTion{\em #1}}

\evensidemargin=\oddsidemargin
\title{\bf Modelling vector-like quarks in partial compositeness framework}

\author{\bf Avik Banerjee\thanks{avik@chalmers.se}\,, Diogo Buarque Franzosi\thanks{diogo.buarque.franzosi@gmail.com}\,, Gabriele Ferretti\thanks{ferretti@chalmers.se}
\\ \medskip 
{}\small\em Department of Physics, Chalmers University of Technology, Fysikg\aa rden, 41296 G\"oteborg, Sweden
}
\date{}

\begin{document}

\maketitle


\begin{abstract}

Composite Higgs models, together with partial compositeness, predict the existence of new scalars and vector-like quarks (partners) at and above the TeV scale. Generically, the presence of these additional scalars opens up new decay topologies for the partners. In this paper we show how to systematically construct the general low energy Lagrangian to capture this feature.  We emphasize the specific pattern in the top-partner spectrum arising in this class of models. We then present a concrete realization in the context of the SU(5)/SO(5) coset. We show that the top-partners in this model can have significant branching ratios to the additional scalars and a third generation quark, compared to the usual Standard Model channels. Amongst the most promising signatures at the LHC are final states containing a diphoton resonance along with a top quark.  
 
\end{abstract}


\bigskip

\newpage
\tableofcontents
\bigskip
\hrule
\bigskip
\section{Introduction}
\label{intro}

The composite Higgs model with partial compositeness (PC) is amongst the few mechanisms available for a symmetry-based explanation of the Electroweak (EW) hierarchy problem in the Standard Model (SM). It realizes the Higgs boson as a pseudo-Nambu-Goldstone boson (pNGB) arising from the breaking of a global symmetry~\cite{Kaplan:1983fs} and uses linear couplings between some SM fermions and composite fermions to generate their masses~\cite{Kaplan:1991dc} and to give the Higgs boson a vacuum expectation value (vev)~\cite{ArkaniHamed:2002qy,Contino:2003ve,Agashe:2004rs}.

If one concentrates on the class of models arising from four-dimensional gauge theories with fermionic matter~\cite{Barnard:2013zea,Ferretti:2013kya} (hyperfermions),
one is led to three minimal types of symmetry breaking:
${\mathcal{G}}/{\mathcal{H}}={\rm SU(5)/SO(5)}$~\cite{Golterman:2015zwa,Ferretti:2014qta,Ferretti:2016upr,Agugliaro:2018vsu}, ${\rm SU(4)/Sp(4)}$~\cite{Barnard:2013zea,Ferretti:2013kya,Cacciapaglia:2014uja,Ferretti:2014qta,Agugliaro:2016clv,Galloway:2016fuo}, $\complexcos$~\cite{Ma:2015gra,Ferretti:2016upr}. We shall refer to these three cosets as the {\it real}, {\it pseudoreal} and {\it complex} case since they arise for hyperfermions in a real, pseudoreal and complex representation of the hypercolor gauge group $G_{\mathrm{HC}}$~\cite{Peskin:1980gc,Preskill:1980mz}. The minimal dimension for the cosets is given by the requirement that it should be possible to obtain the Higgs field as a bidoublet $H\in ({\mathbf{2}},{\mathbf{2}})$ of ${\rm SU}(2)_L\times {\rm SU}(2)_R \subset {\mathcal{H}}$~\cite{Mrazek:2011iu}. Note that the minimal composite Higgs model ${\rm SO}(5)/{\rm SO}(4)$~\cite{Contino:2003ve,Agashe:2004rs,Contino:2006qr,Panico:2011pw,Contino:2011np,Azatov:2011qy,DeSimone:2012fs,Matsedonskyi:2012ym,Marzocca:2012zn,Pomarol:2012qf,Panico:2012uw,Carena:2014ria,Montull:2013mla,Carmona:2014iwa,Panico:2015jxa,Niehoff:2015iaa,Kanemura:2016tan,Gavela:2016vte,Banerjee:2017wmg,Liu:2017dsz}  does not belong to this class, but the next-to-minimal one ${\rm SO}(6)/{\rm SO}(5)\equiv {\rm SU}(4)/{\rm Sp}(4)$~\cite{Gripaios:2009pe,Redi:2012ha,Serra:2015xfa,Low:2015qep,Cai:2015bss,Arbey:2015exa,Niehoff:2016zso,Golterman:2017vdj,Alanne:2018wtp,Murnane:2018ynd,Bian:2019kmg,DeCurtis:2019rxl,Frigerio:2012uc,Marzocca:2014msa,Fonseca:2015gva,Kim:2016jbz,Espinosa:2011eu,Banerjee:2017qod,Erdmenger:2020lvq} does. The low energy theory at the multi-TeV scale comprises the $G_{\mathrm{HC}}$ invariant bound states of these hyperfermions with different spins and SM quantum numbers. Bilinears of the hyperfermions lead to pNGB scalars and spin-1 vector resonances in the infrared (IR), while trilinears give rise to vector-like fermionic bound states (commonly called vector-like quarks or fermionic partners).

We first present the structure of the IR lagrangian describing the interactions of the third family of quarks and the SM gauge bosons with the fermionic partners and pNGBs of the strongly coupled sector.
This is needed since the spectra and the couplings of these theories are highly non-generic, and thus simplified models may miss interesting signatures for searches or falsely point to some that cannot be realized in the full theory. Also, tools are being developed to automatize the simulation of general models of PC, with future hadron colliders in mind.
Extending the field content without guidance from symmetry principles leads to many undetermined parameters. However, once the symmetries of the underlying theory are employed, the number of these parameters is greatly reduced. 

We use the PC paradigm to construct the low energy Lagrangian of the composite sector and present it for the relevant cosets. We further chalk out the steps to extract the interactions of the elementary fields with the composite ones, and provide details in the appendix. The mass spectrum of the fermionic partners in the PC framework has some specific patterns, independent of the choice of coset. In particular, a group of nearly degenerate fermionic bound states is expected in this class of models.   

We then consider a specific realization of the real ($\rm SU(5)/SO(5)$) case and survey its experimental signatures. The choice of focusing on this coset is partly motivated by the recent interest in searching for exotic signatures of top-partner decays $t'\to t\, (S\to \gamma\, \gamma)$ in the notation of~\cite{Benbrik:2019zdp}, (see also \cite{Banerjee:2016wls,Wang:2020ips}), which are more easily realized in this coset. It follows that the branching ratios to the SM final states such as $th$, $tZ$ and $bW$ are reduced compared to those into pNGBs and third generation quarks. In particular, we analyze a specific scenario where the pair production of top-partners, with one of them decaying into a $t\gamma\gamma$ final state, has a cross section of the order of a few femtobarn. 

Our focus is on the production and decay of composite fermions and EW pNGBs, but it should be noticed that the bound states arising from underlying theories of this kind include additional types of composite particles. Some of these additional particles have been studied elsewhere: colored pNGBs~\cite{Cacciapaglia:2015eqa,Belyaev:2016ftv}, fermions in non-triplet irreps of color~\cite{Cacciapaglia:2021uqh}, vector resonances~\cite{Azatov:2015xqa,BuarqueFranzosi:2016ooy,Yepes:2018dlw,Dasgupta:2019yjm,Banerjee:2021efl}, and axion-like particles~\cite{Cacciapaglia:2019bqz,BuarqueFranzosi:2021kky}. One can also envisage a variety of additional decay channels of the fermionic partners as discussed in \cite{Bizot:2018tds,Cacciapaglia:2019zmj,Xie:2019gya}, see also \cite{Matsedonskyi:2014lla,Corcella:2021mdl,Dasgupta:2021fzw,Chala:2017xgc,Ramos:2019qqa}.

The paper is structured as follows. Sections~\ref{lagrangian} and \ref{massdiag} together with appendix \ref{IR_theory} present the general construction of the models while section~\ref{pheno} together with appendices \ref{IR_realcase_theory} and \ref{decaywidths} deals with the specific $\rm SU(5)/SO(5)$ coset and its diphoton signal. We offer our conclusions in section~\ref{concl}.

\section{IR Lagrangian}
\label{lagrangian}

The purpose of this section is to present the various components of the IR Lagrangian describing the interactions of the composite sector with the SM vector bosons and the quarks of the third generation. At this stage we keep the presentation general, including all the minimal cosets arising in these underlying models, while the main interest of the later phenomenological section (section~\ref{pheno} and appendix~\ref{IR_realcase_theory}) is in the $\rm SU(5)/SO(5)$ coset.

We split the Lagrangian into several parts for ease of discussion
\begin{equation}
\label{L_tot}
{\mathcal{L}}={\mathcal{L}}_{\mathrm{elem}}+{\mathcal{L}}_{\mathrm{comp}}\,,\quad {\rm with} \quad {\mathcal{L}}_{\mathrm{comp}}={\mathcal{L}}_{\mathrm{pNGB}}+{\mathcal{L}}_{\mathrm{anom}}+{\mathcal{L}}_{\Psi^2}+{\mathcal{L}}_{\mathrm{PC}}-V_{\mathrm{pot}}\,.
\end{equation}
The elementary Lagrangian $\mathcal{L}_{\rm elem}$ for the third generation quarks and vector bosons is identical to the SM Lagrangian
\begin{equation}
\label{L_elem}
\mathcal{L}_{\rm elem}=\bar{q}_Li\slashed D q_L+\bar{t}_Ri\slashed D t_R+\bar{b}_Ri\slashed D b_R-\frac{1}{4}G^a_{\mu\nu}G^{a\mu\nu}-\frac{1}{4}W^a_{\mu\nu}W^{a\mu\nu}-\frac{1}{4}B_{\mu\nu}B^{\mu\nu}\,.
\end{equation}
In what follows we discuss each part of the composite Lagrangian $\mathcal{L}_{\mathrm{comp}}$.  More details are given in appendix\,\ref{IR_theory}. In this work, for simplicity we assume CP invariance and all couplings to be real.

\subsection{The pNGB kinetic term}
\label{kinetic}

The canonically normalized electroweak pNGBs $\Pi=\pi_i\hat{T}^i$ can be written in terms of the broken generators $\hat{T}^i$ as a matrix $\Sigma\equiv\Omega(\theta) \exp\left(i\beta\Pi/f\right)$ transforming non linearly~\cite{Coleman:1969sm,Callan:1969sn} under a global transformation $g\in \mathcal{G}$ as $\Sigma\to g\Sigma h^{-1}(\Pi,g)$ where $h(\Pi,g)\in \mathcal{H}$. Here $f\sim$ TeV represents the pNGB decay constant, and following the conventions in~\cite{Ferretti:2016upr} $\beta=1,\sqrt{2},\sqrt{2}$ for the real, pseudoreal and complex case respectively. 

The explicit pNGB content are given in appendix\,\ref{sec_pngbs} for the three cosets under consideration.  
Throughout this paper we assume that the vacuum misalignment, leading to EW symmetry breaking (EWSB), is caused by a nonzero vev of the Higgs doublet alone, while other pNGB scalars do not receive any vevs. 
The true vacuum after EWSB can be obtained by exponentiating the Higgs vev and is denoted by a matrix $\Omega(\theta)$ as in~\cite{Ferretti:2016upr} where the angle $\theta$ parameterizes the misalignment of the vacuum. 

For the real and pseudoreal cases one can also construct a matrix $U\equiv\Sigma\epsilon\Sigma^T$  which transforms as $U\to gUg^T$. Here $\epsilon$ represents a symmetric (antisymmetric)  $\mathcal{H}$ invariant tensor $h\epsilon h^T=\epsilon$ for the real (pseudoreal) case respectively. For the complex case it is more convenient to split the elements of $\rm SU(4)_l\times SU(4)_r$ as $g=(g_l,g_r)$, let $\Sigma_l\to g_l\Sigma_l h^{-1}$, $\Sigma_r\to g_r\Sigma_r h^{-1}$, and define $U=\Sigma_l\Sigma_r^\dagger\to g_lUg_r^\dagger$, with $h\in {\rm SU(4)_d}$. Our explicit formulas will be mostly based on the real/pseudoreal case with the modification for the complex case left understood\footnote{To avoid confusion we use the lowercase letters $l,r$ to denote the two simple $\rm SU(4)$ factors of $\mathcal G$ in the complex case and reserve the uppercase $L,R$ to the $\rm SU(2)$ factors in $\mathcal H$ and their related chiral structure.}.

While $\Sigma$ is required to define the interactions of the vector-like quarks with the SM fermions through PC, the self interactions and gauge interactions of the pNGBs can be expressed more conveniently using $U$. At $\mathcal{O}(p^2)$, the kinetic term of the pNGBs can be written using the matrix $U$ as 
\begin{equation}
\label{L_kin}
\mathcal{L}_{\rm pNGB}=\frac{f^2\beta^2}{16}{\rm tr}\left[(D_\mu U)^\dagger (D^\mu U)\right]\,.
\end{equation}
We introduce $\beta$ in order to simultaneously canonically normalize the kinetic terms of the pNGBs and ensure that the masses of $W$ and $Z$ bosons can be written uniformly as
\begin{equation}
\label{gauge_mass}
M_W^2=M_Z^2c_W^2=\frac{g^2}{4}f^2s_\theta^2\,.
\end{equation}
Here $c_W\equiv\cos\theta_W$ denotes the cosine of the weak mixing angle, $s_\theta\equiv\sin\theta$ and the definition of electroweak scale is fixed as $v=fs_\theta=246$ GeV. Note that the tree level relation $\rho=1$ is preserved since we assume that the vevs of all other non-standard pNGB scalars are zero. The interactions of the weak gauge bosons with the 125 GeV Higgs boson can thus be written as
\begin{equation}
\label{hVV_int}
\mathcal{L}=M_W^2\left(2 c_\theta \frac{h}{v}+c_{2\theta}  \frac{h^2}{v^2}+...\right)\left[W_\mu^+W^{-\mu}+\frac{1}{2c_W^2}Z_\mu Z^\mu\right]\,. 
\end{equation}
The $hVV$ and $hhVV$ couplings ($V=W,Z$) are modified with respect to the SM by a universal factor $c_\theta$ and $c_{2\theta}$ respectively for any compact coset~\cite{Liu:2018qtb}. The Lagrangian Eq.\,\eqref{L_kin} also describes the interactions of one or two EW gauge bosons with \emph{pairs} of additional (non-Higgs) pNGBs, generically denoted as $\pi$, via the covariant derivative \cite{Ferretti:2016upr}. Note that $\pi VV$ terms are absent because of the absence of a vev for $\pi$.

\subsection{The anomalous terms}

In the absence of a vev for $\pi$, the interactions of a single pNGB with the gauge bosons are given by the hyperquark anomaly, described by a Wess-Zumino-Witten term (WZW).
The WZW terms involving one pNGB and two gauge bosons can be written in a coset independent way in terms of differential forms as~\cite{Witten:1983tw, Kaymakcalan:1983qq}
\begin{align}
\nonumber
S_{\mathrm WZW} \supset \frac{i \dim(\psi)}{48\pi^2}\int & \tr\bigg(dA A dU U^\dagger + A dA dU U^\dagger + dA^\prime A^\prime U^\dagger dU \\
& + A^\prime dA^\prime U^\dagger dU - dA dU A^\prime U^\dagger + dA^\prime dU^\dagger A U \bigg),
\end{align}
 where $\dim(\psi)$ is the dimension of the hypercolor irrep of the hyperfermion $\psi$ giving rise to the EW coset, and $A$ denotes the Lie algebra valued one form $A\equiv(g W_\mu^aT^a_L+g^\prime B_\mu T^3_R)dx^\mu$.
For the $\rm SU(4)/Sp(4)$ and $\rm SU(5)/SO(5)$ \cite{Ferretti:2016upr} cosets $A^\prime=-A^T$ while for the $\complexcos$ \cite{Ma:2015gra} case $A^\prime=A$. After integrating by parts and expanding to leading order in the pNGBs the anomaly Lagrangian $\mathcal{L}_{\rm anom}$ can be written in terms of the physical gauge fields  as
\begin{align}
\nonumber
\mathcal{L}_{\rm anom} & = \frac{e^2\dim(\psi)}{48\pi^2 f} \bigg[\sum_i \pi^0_i\left(K^i_{\gamma\gamma}F_{\mu\nu}\tilde{F}^{\mu\nu}+K^i_{\gamma Z}F_{\mu\nu}\tilde{Z}^{\mu\nu}+K^i_{ZZ}Z_{\mu\nu}\tilde{Z}^{\mu\nu}+K^i_{WW}W^+_{\mu\nu}\tilde{W}^{-\mu\nu}\right)\\
&+\sum_j \pi^+_j \left( K^j_{\gamma W}F_{\mu\nu}\tilde{W}^{-\mu\nu}+K^j_{ZW}Z_{\mu\nu}\tilde{W}^{-\mu\nu}\right)+{\rm h.c.}+\sum_k  \pi^{++}_k K^k_{W^-W^-}W^-_{\mu\nu}\tilde{W}^{-\mu\nu}+{\rm h.c.}\bigg]. \label{WZWterms}
\end{align}
Here $\pi^0_i$, $\pi^\pm_j$ and $\pi^{\pm\pm}_k$ represent any pNGBs for a generic coset with electric charge $Q=0,\pm1$ and $\pm2$, respectively. In Table\,\ref{WZW_coups} we list the coefficients of the $\mathcal{L}_{\rm anom}$ for different pNGBS in the custodial basis for the three minimal cosets (see appendix\,\ref{sec_pngbs} for notations). Note that for the $\complexcos$ coset only $\eta$ couples to the anomaly term and for both $\rm SU(4)/Sp(4)$ as well as $\complexcos$ cosets $K^i_{\gamma\gamma}=0$, as already pointed out in \cite{Ma:2015gra}. Also, for the real case $\chi_3^0$ does not appear in $\mathcal{L}_{\rm anom}$.
\begin{table}[t!]
	\begin{center}
		\begin{tabular}{cccccccccc}
			\hline\hline
			Coset & $\pi_i$ & $K^i_{\gamma\gamma}$ & $K^i_{\gamma Z}$ & $K^i_{ZZ}$ & $K^i_{WW}$ & $K^i_{\gamma W}$ & $K^i_{ZW}$ & $K^i_{W^-W^-}$\\
			\hline
			
			\multirow{1}{*}{$\rm \frac{SU(4)}{Sp(4)}$} & $\eta$ & $0$ & $\frac{3 c_\theta}{s_{W}c_W}$ & $\frac{3c_{2W}c_\theta}{2 s_{W}^2c_W^2}$ & $\frac{3c_\theta}{s_{W}^2}$ & -- & -- & --\\
			\hline
			
			\multirow{7}{*}{$\rm \frac{SU(5)}{SO(5)}$} & $\chi^0_5$ & $-2\sqrt{3}$ & $-\frac{4\sqrt{3}c_{2W}}{s_{2W}}$ & $\frac{(1-3c_{4W}+2c_{2\theta})}{\sqrt{3}s_{2W}^2}$ & $\frac{s^2_{\theta}}{\sqrt{3}s_{W}^2}$ & -- & -- & --\\
			
			& $\chi^0_1$ & $\sqrt{6}$ & $\frac{2\sqrt{6}c_{2W}}{s_{2W}}$ & $\frac{(1+6c_{4W}-7c_{2\theta})}{2\sqrt{6}s_{2W}^2}$ & $\frac{7}{2\sqrt{6}}\frac{s^2_{\theta}}{s_{W}^2}$ & -- & -- & --\\
			
			& $\eta$ & $3\sqrt{\frac{2}{5}}$ & $\sqrt{\frac{2}{5}}\frac{6c_{2W}}{s_{2W}}$ & $\frac{3(3c_\theta^2+c_{4W})}{4\sqrt{10}c_W^2s_W^2}$ & $\frac{3(3c_{2\theta}+5)}{4\sqrt{10}s_W^2}$ & -- & -- & -- \\
			
			& $\chi^+_3$ & -- & -- & -- & -- & $-\frac{3c_\theta}{s_W}$ & $\frac{3c_\theta}{c_W}$ & -- \\
			
			& $\chi^+_5$ & -- & -- & -- & -- & $\frac{3i}{s_W}$ & $\frac{i(3c_{2W}-c_{2\theta}-2)}{2c_Ws_W^2}$ & -- \\
			
			& $\chi^{++}_5$ & -- & -- & -- & -- & -- & -- & $-\frac{s_\theta^2}{\sqrt{2}s_W^2}$ \\
			\hline
			
			\multirow{1}{*}{$\rm \frac{SU(4)_l\times SU(4)_r}{SU(4)_d}$} & $\eta$ & $0$ & $\frac{3 c_\theta}{s_{W}c_W}$ & $\frac{3c_{2W}c_\theta}{2 s_{W}^2c_W^2}$ & $\frac{3 c_\theta}{s_{W}^2}$ & -- & -- & -- \\
			\hline\hline
	\end{tabular}
\caption{\small\it Coefficients of the anomaly terms in Eq.\,\eqref{WZWterms} uniformly normalized by a factor $\frac{e^2\dim(\psi)}{48\pi^2 f}$. Here $\eta$ always denotes a SM singlet while the remaining coefficients are expressed in the custodial basis but otherwise agree with those in~\cite{Ferretti:2016upr}. The pNGBs not appearing in the table do not couple to the anomaly terms.}
\label{WZW_coups}
\end{center}	
\end{table}

\subsection{The Lagrangian for vector-like quarks}
\label{Ltpartner}

Vector-like fermionic partners ($\Psi$) are built out of $G_{\mathrm{HC}}$-invariant trilinears involving two types of hyperfermions $\psi$ and $\chi$.
The $\rm SU(3)_c$ quantum number of the partner is carried by the $\chi$-type hyperfermions, which are however not charged under $\mathcal{G}$. On the other hand, $\psi$ transforms as a fundamental ($F$) of $\mathcal{G}$~\footnote{We use $N, F, A, S, D$ to represent the si{\it N}glet, {\it F}undamental, {\it A}ntisymmetric, {\it S}ymmetric, and a{\it D}joint irreps of both $\mathcal{G}$ and $\mathcal{H}$, which one is being used should be evident from the context. For complex coset the $\psi$ transforms as $(F,1)+(1,\bar{F})$. In order to simplify the discussion we will often refer to both $F$ and $\bar{F}$ as ``fundamental'' in the text.}.
The trilinear composite operators can thus be divided into two major categories of the type $\chi\psi\chi$ (one-index irrep of $\mathcal{G}$) and $\psi\chi\psi$ (two-index irrep of $\mathcal{G}$) respectively.

Below the $\mathcal{G}\to \mathcal{H}$ symmetry breaking scale the irreps of $\mathcal{G}$ should be decomposed under the unbroken global $\mathcal{H}$.
This implies that $\chi\psi\chi$-type partners transform as the irreps in the decomposition of $F$ of $\mathcal{G}$ on restriction to $\mathcal{H}$, while the $\psi\chi\psi$-type partners belong to the irreps in the decomposition of $F\times F$. Thus, in matrix notation, the top-partners transform as $\Psi_F\to h\Psi_F$ (for fundamental of $\mathcal{H}$), $\Psi_{S,A}\to h\Psi_{S,A} h^T$ (for symmetric or anti-symmetric of $\mathcal{H}$) and  $\Psi_D\to h\Psi_D h^\dagger$ (for adjoint of $\mathcal{H}$ in the complex case). 

Taking into account these informations from the UV, we consider a single irrep (up to two-index) of $\mathcal{H}$ for the fermionic partner as the relevant dynamic degree of freedom. For any coset, it is convenient to decompose $\Psi$ in terms of $\rm SU(2)_L\times SU(2)_R \subset {\mathcal{H}}$. 
Explicit matrices for $\Psi$ in different irreps are given in the appendix\,\ref{top_partners} for the three minimal cosets. The Lagrangian for the fermionic partners is then given by\footnote{For the one-index irreps the trace should be interpreted as the usual invariant product.}
\begin{equation}
\label{L_tp}
\mathcal{L}_{\Psi^2}={\rm tr}\left[\bar{\Psi} i\slashed D \Psi\right]-M{\rm tr}\left[\bar{\Psi} \Psi\right]+\kappa{\rm tr}\left[\bar{\Psi}\slashed d \Psi\right],
\end{equation}
where the covariant derivative is
\begin{align}
\label{tp_covder}
D_\mu \Psi= \partial_\mu  \Psi-i v_\mu  \Psi- i X e\left(A_\mu -\frac{s_W}{c_W} Z_\mu \right)\Psi-ig_s G^a_\mu\frac{\lambda^a}{2} \Psi\,.
\end{align}
In the second term of the above equation $v_\mu$ acts on $\Psi$ in the appropriate representation. The third term in Eq.\eqref{tp_covder} corresponds to the interactions along the  additional factor of $\rm U(1)_X$ that is the minimal additional gauge degree of freedom necessary to reproduce the correct hypercharge (given by $Y=T^3_R+X$) of the SM quarks\footnote{Note that the pNGBs are uncharged under the $\rm U(1)_X$.}. The matrix-valued $d_\mu$ and $v_\mu$ symbols can be calculated using the CCWZ formalism~\cite{Coleman:1969sm,Callan:1969sn} and are given by following expressions,
\begin{align}
d_\mu=\sum_{i=1}^{{\small\mathrm{dim}(\mathcal{G}/\mathcal{H})}}\hat{T}^i{\rm tr}\left[\hat{T}^i\Sigma^{-1}\left(i\partial_\mu\Sigma+eV_\mu\Sigma\right)\right]\,,\quad
v_\mu=\sum_{a=1}^{{\small\mathrm{dim}(\mathcal{H})}}T^a{\rm tr}\left[T^a\Sigma^{-1}\left(i\partial_\mu\Sigma+eV_\mu\Sigma\right)\right]\,,
\end{align}
where $\hat{T}^i$ ($T^a$) denotes the broken (unbroken) generators of $\mathcal{G}$ and $V_\mu$ is given by
\begin{align}
\label{Vmu}
V_\mu=\frac{W_\mu^+}{s_W}\frac{T^1_L+iT^2_L}{\sqrt{2}}+\frac{W_\mu^-}{s_W}\frac{T^1_L-iT^2_L}{\sqrt{2}}+\left(A_\mu+\frac{c_W}{s_W}Z_\mu\right)T^3_L+\left(A_\mu-\frac{s_W}{c_W}Z_\mu\right)T^3_R\,.
\end{align}
The term proportional to $\kappa$ in Eq.\,\eqref{L_tp} leads to the derivative interactions of the partners with the pNGB fields and belongs exclusively to the strong sector.

\subsection{The partial compositeness Lagrangian}

The PC mechanism relies on the linear mixing between the top quark and the top-partner which explicitly breaks the global symmetry of the strong sector. In order to parameterize this explicit breaking by coupling the SM third generation quarks to the top-partners and the pNGBs, we use spurionic embeddings of the SM fermions into the irreps of $\mathcal{G}$. 

For the real and pseudoreal cases we consider spurions which transform as $N, F, A, S,$ and $D$ of $\mathcal{G}$ (for the explanation of the notation, see section\,\ref{Ltpartner}). 
For the complex case the spurions are classified according to the representations $(\rho_l, \rho_r)$ where $\rho_{l,r}=N, F\dots D$, with the additional possibility of a {\it B}ifundamental $B$, but the idea behind the construction is the same. (See~\cite{Ferretti:2016upr,Alanne:2018wtp} for more details.)

The transformation properties of the spurions under a global  $g\in\mathcal{G}$ are given, in matrix notation, by
\begin{equation}
\label{spurion_transform}
N\to N\,, \quad F\to g F\,, \quad A \to g A g^T\,, \quad S\to g S g^T\,, \quad D\to g D g^\dagger\,.
\end{equation} 
The left-handed quark doublet ($q_L$) should be embedded into a $(\mathbf{2},\mathbf{2})_{2/3}$ of $\rm SU(2)_L\times SU(2)_R\times U(1)_X$, while for the right-handed top ($t_R$) either a $(\mathbf{1},\mathbf{1})_{2/3}$, or the $T^3_R=0$ component of a $(\mathbf{1},\mathbf{3})_{2/3}$ can be used. The embedding of $q_L$ in $(\mathbf{2},\mathbf{2})$ ensures that the corrections to the $Z\to b_L\bar{b}_L$ decay width is under control due to the custodial protection~\cite{Agashe:2006at}. 
The explicit embedding matrices for the pseudoreal, real and complex cases are given in appendix\,\ref{spurions}. 

The structure of the formally $\mathcal{G}$ invariant PC Lagrangian is given as
\begin{table}[t!]
	\begin{center}
		\begin{tabular}{cccccccc}
			\hline\hline
			Spurions & $\Psi_N$ & $\Psi_F$ & $\Psi_{A}$ & $\Psi_{S}$ \\
			\hline
			$N$ & $\Psi_{N}$ & $\times$ & 0 & 0 \\
			$F$ & $\times$ & $F^\dagger\Sigma\Psi_{F}$ & $\times$ & $\times$ \\
			$A$ & ${\rm tr}\left[A^\dagger\Sigma\epsilon\Sigma^T\right]\Psi_{N}$ & $\times$ & ${\rm tr}\left[A^\dagger\Sigma\Psi_{A}\Sigma^T\right]$ & 0 \\
			$S$ & ${\rm tr}\left[S^\dagger\Sigma\epsilon\Sigma^T\right]\Psi_{N}$ & $\times$ & 0 & ${\rm tr}\left[S^\dagger\Sigma\Psi_{S}\Sigma^T\right]$ \\
			$D$ & 0 & $\times$ & ${\rm tr}\left[D^\dagger\Sigma\Psi_{A}\epsilon\Sigma^\dagger\right]$ & ${\rm tr}\left[D^\dagger\Sigma\Psi_{S}\epsilon\Sigma^\dagger\right]$ \\
			\hline\hline
	\end{tabular}
\caption{\small\it Formally $\mathcal{G}$ invariant operators at leading order giving rise to PC of the SM quarks. The $\times$ symbol implies no possible invariant can be constructed while the zeros denote that the operators vanish identically due to symmetry properties of the spurions and the irreps of the fermionic partners. For the real (pseudoreal) case the expression ${\rm tr}\left[A^\dagger\Sigma\epsilon\Sigma^T\right]$, (${\rm tr}\left[S^\dagger\Sigma\epsilon\Sigma^T\right]$) vanishes, since $\epsilon$ is symmetric (antisymmetric).}
\label{invariants}
\end{center}	
\end{table}
\begin{equation}
\label{L_PC}
\mathcal{L}_{\rm PC}=y_L f \bar{\hat{q}}_L\mathcal{O}_R + y_R f \bar{\hat{t}}_R\mathcal{O}_L+{\rm h.c.}
\end{equation}
Here $\hat{q}_L$ and $\hat{t}_R$ denote the embedding of the elementary quarks into the incomplete $\mathcal{G}$ multiplets above. For example, if both $q_L$ and $t_R$ are embedded in the adjoint irrep of $\mathcal{G}$, we can write $\hat{q}_L=t_L D_{t_L}+b_L D_{b_L}$ and $\hat{t}_R=t_R D_{t_R}$. To construct the invariant operators $\mathcal{O}_{L,R}$, the partner $\Psi$ is dressed with appropriate insertions of the pNGB matrix $\Sigma$. In Table\,\ref{invariants} we present the leading order invariant terms of Eq.\,\eqref{L_PC} for the real and pseudoreal case. The construction is the same in the real and pseudoreal case if one uses the appropriate invariant tensor $\epsilon$ (symmetric or anti-symmetric). 

For the complex case one simply needs to keep track of the difference between the left and right factors of $\mathcal{G}$. For instance, one can have the following two invariants with antisymmetric tensors: 
${\rm tr}\left[A_r^\dagger\Sigma_r\Psi_{A}\Sigma_r^T\right]$ and ${\rm tr}\left[A_l^\dagger\Sigma_l\Psi_{A}\Sigma_l^T\right]$ where $A_l\to g_l A_l g_l^T$ or $A_r\to g_r A_r g_r^T$ and similarly for the adjoint and symmetric. In the complex case one can also use the bifundamental $B\to g_l B g_r^T$ and similar for various combinations of fundamentals and anti-fundamentals e.g. $B'\to g_l B' g_r^\dagger$. Here the invariants are simply ${\rm tr}\left[B^\dagger\Sigma_l\Psi_{A}\Sigma_r^T\right]$ or ${\rm tr}\left[B^\dagger\Sigma_l\Psi_{S}\Sigma_r^T\right]$ in the first case and ${\rm tr}\left[B^{\prime\dagger}\Sigma_l\Psi_N\Sigma_r^\dagger\right]$ or 
${\rm tr}\left[B^{\prime\dagger}\Sigma_l\Psi_{D}\Sigma_r^\dagger\right]$ in the second case.

\subsection{The scalar potential}

We have now come to the last term in Eq.\,\eqref{L_tot}, namely the scalar potential. This term is the most model dependent and prone to computational difficulties since it is fully generated by the terms explicitly breaking the global symmetry $\mathcal{H}$. 
The contributions of the symmetry breaking interactions to the potential can be split as 
\begin{equation}
V_{\rm pot}=V_m+V_g+V_t\,,   \label{fullpot}
\end{equation}
where $V_m$ denotes the contribution from the bare hyperfermion mass term which can be written as\footnote{This is not the most general expression. Any matrix preserving custodial symmetry can be used in place of $\epsilon$.}
\begin{equation}
\label{hyper_pot}
V_m= B_m f^4 {\rm tr}\left[\epsilon^* U + \epsilon U^*\right] + B_m^\prime f^4 {\rm tr}\left[(\epsilon^* U)^2+(\epsilon U^*)^2\right]\,.
\end{equation}
The coefficients $B_m, B_m^\prime$ are dimensionless low energy parameter encoding the strong dynamics. We have included a $B^\prime_m$ contribution, in the same spirit as \cite{Knecht:1994zb,Knecht:1994wf,Knecht:1995ja}. Although $f$ should be thought of as an inverse coupling and inserted according to the rules of naive dimensional analysis, we chose to use it as the only dimensionful quantity for convenience, hence the unusual $f$ dependence.

Similarly one-loop contributions from the gauge bosons $V_g$ are given by
\begin{equation}
\label{gauge_pot}
V_g=B_g f^4 {\rm tr}\left[g^2T^a_L UT^{a*}_LU^\dagger+g^{\prime 2}T^3_R UT^{3*}_RU^\dagger\right]\,,
\end{equation}
for the real and pseudoreal cases, while for the complex case both $T^{a*}_L$ and $T^{3*}_R$ should be replaced by $-T^{a}_L$ and $-T^{3}_R$, respectively. 

The top quark contribution to the potential  $V_t$ depends on the specific spurionic representations in which $q_L$ and $t_R$ are embedded. 
Generically, the lowest order invariants are formed using the spurions in the various representations and are given by terms like 
\begin{equation}
\label{spurion_pot}
(F^\dagger U F^*)(F^T U^* F)\,, \quad {\rm tr}(AU^*){\rm tr}(A^*U)\,, \quad {\rm tr}(SU^*){\rm tr}(S^*U)\,, \quad \mbox{and} \quad {\rm tr}(DUD^*U^*)\,,
\end{equation}
for the real and pseudoreal case, and similarly for the complex case. 
Note that the antisymmetric spurion does not contribute to the potential in the real case and the symmetric one does not contribute in the pseudoreal case. Also, none of the irreps $(\mathbf{1},\rho)$ or $(\rho,\mathbf{1})$ contribute in the complex case.

Writing the full potential as in Eq.\,\eqref{fullpot} one proceeds first by selecting those spurions that guarantee the absence of tadpoles for the non-Higgs pNGBs. For these selected representations one then fixes two linear combinations of the low energy coefficients $B$ by imposing the correct Higgs mass and vev. The remaining coefficients are then varied in the stability region of the potential to read off the spectrum of pNGBs.
Detailed discussions about the scalar potential for the various cosets can be found in~\cite{Gripaios:2009pe,Ma:2015gra,Ferretti:2016upr,Agugliaro:2018vsu}.

\section{Spectrum of the fermionic partners}
\label{massdiag}

Having constructed the full Lagrangian in section~\ref{lagrangian}, we can now study the generic properties of the fermionic spectrum arising in these models. Here we discuss the classical mass matrices arising from combining the Dirac mass in Eq.\,\eqref{L_tp} and the contribution of Eq.\,\eqref{L_PC} after EWSB. The only partners involved in the quadratic part of Eq.\,\eqref{L_PC} are those with the same quantum number as the top or bottom quarks; the other exotic partners are unaffected by PC and remain degenerate with tree level mass set by $M$ in Eq.\,\eqref{L_tp}. We thus focus our attention on these two sectors. The generic structure of their mass matrix is that of a $n\times n$ matrix
\begin{equation}
\label{mass_matrix}
\mathcal{M}_{2/3}=\left(\begin{array}{@{}c|c@{}}
0
& \omega^t_L(\theta)^T \\
\hline
\omega^t_R(\theta) &
M \mathbb{1}_{n-1}
\end{array}\right)\,, \quad
\mathcal{M}_{-1/3}=\left(\begin{array}{@{}c|c@{}}
0
& \omega^b_L(\theta)^T \\
\hline
\omega^b_R(\theta) &
M \mathbb{1}_{n-1}
\end{array}\right)\,,
\end{equation} 
where $\omega^{t,b}_{L,R}(\theta)$ denote $(n-1)$ dimensional vectors capturing the mixing between the elementary and composite fermions in Eq.\,\eqref{L_PC}. In the case where a $b_R$ partner is absent, we have $\omega^b_R(\theta)=0$ and the bottom quark cannot acquire a mass this way. In this case one needs to resort to the usual bilinear couplings to give a mass to the bottom quark, expressed by replacing the upper left zero-entry of $\mathcal{M}_{-1/3}$ by a non-zero parameter. The elements of $\omega^{t,b}_{L,R}(\theta)$ are proportional to $y_{L,R}f$ and trigonometric functions of $\theta$. 

\subsubsection*{Singular value decomposition of $\mathcal{M}_{2/3}$:}

\begin{figure}[t!]
	\centering
	\includegraphics[width=0.455\textwidth,keepaspectratio]{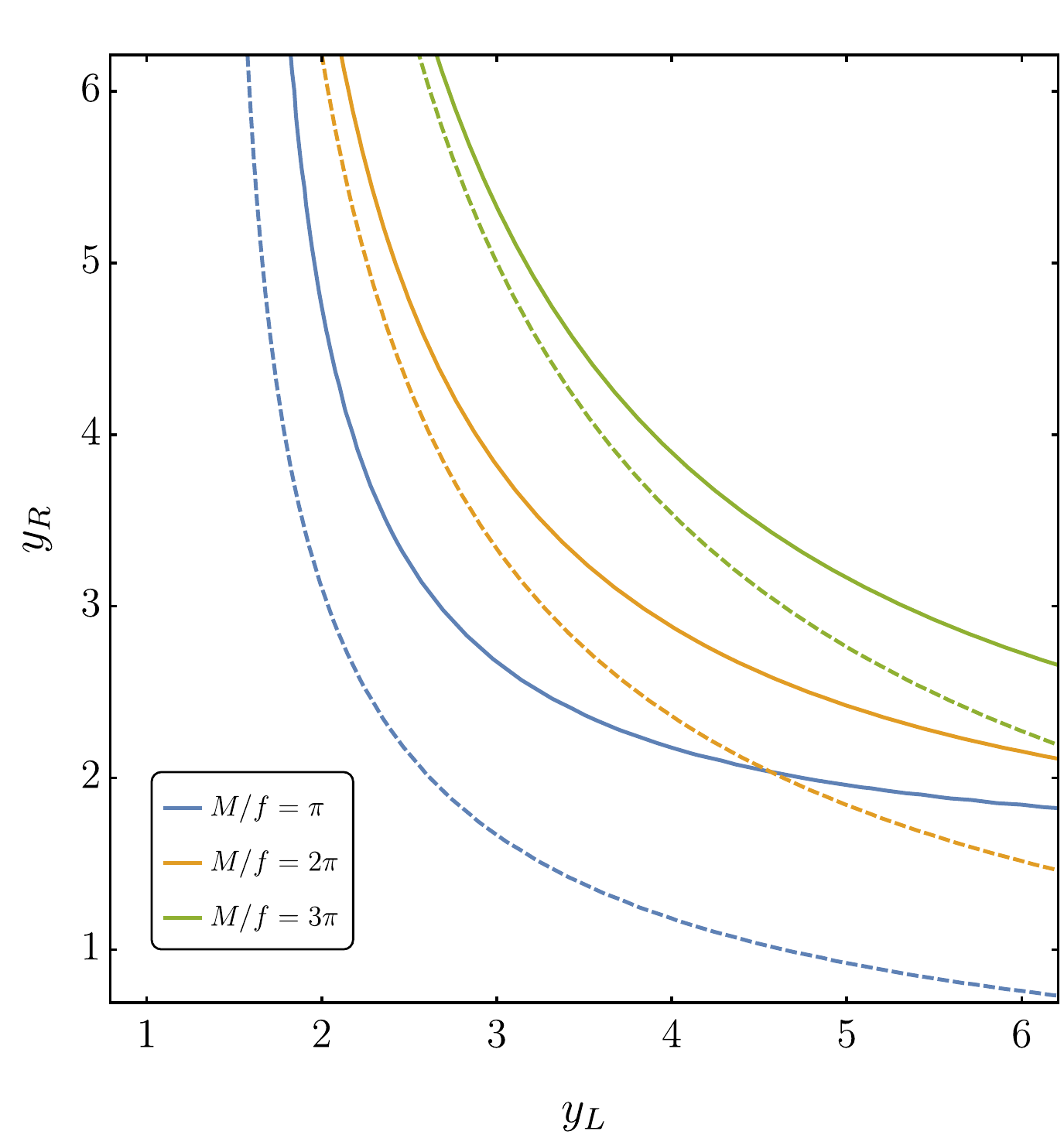}
	\hspace{0.2cm}
	\includegraphics[width=0.48\textwidth,keepaspectratio]{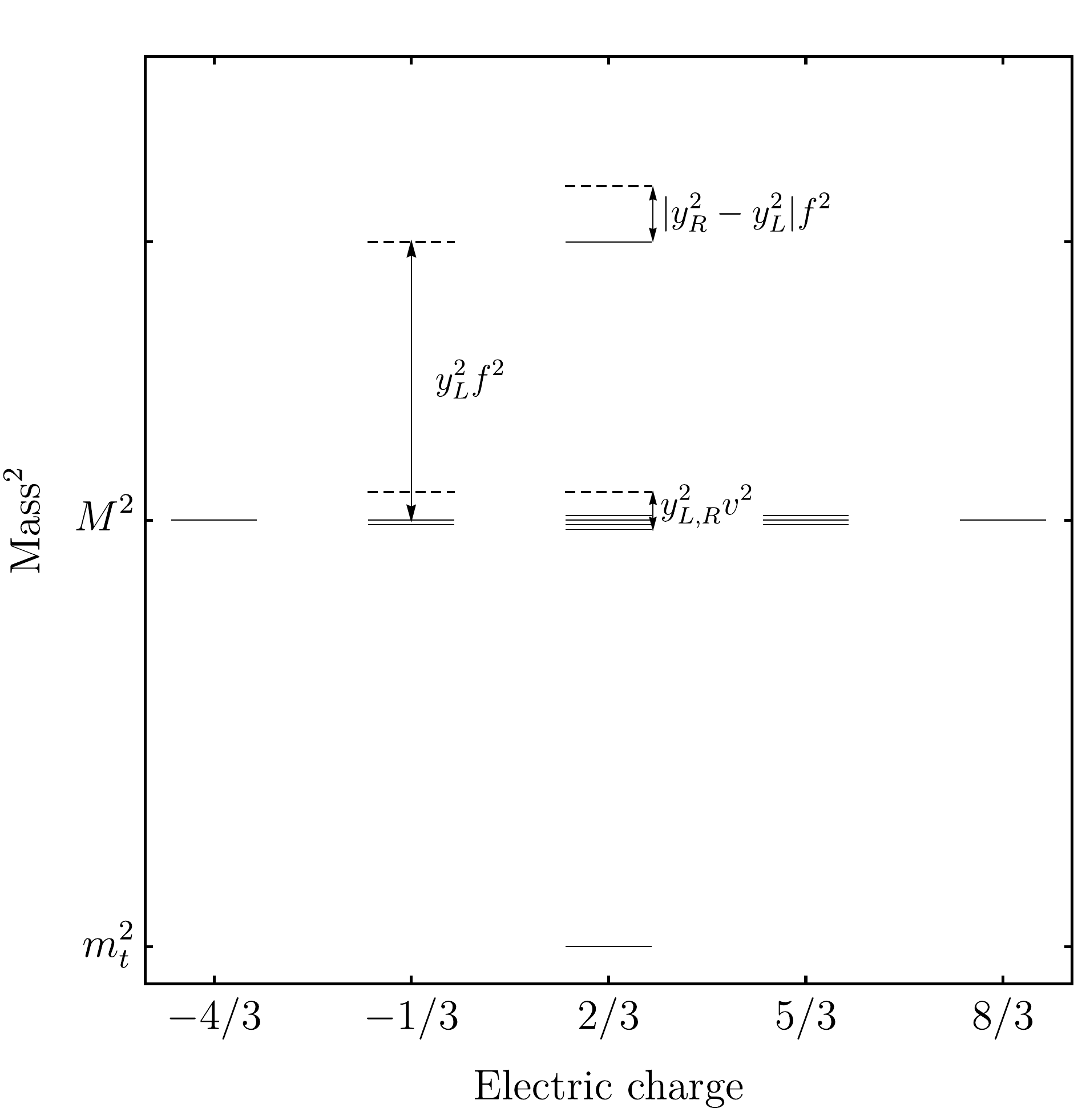}
	\caption{\small\it Left: The $m_t=173$ GeV contours in the $y_R-y_L$ plane for $M/f=\pi$ (blue), $2\pi$ (yellow) and $3\pi$ (green). The solid (dashed) contours refer to the models of Type  I (II), as given in Eq.\,\eqref{top_mass}. Right: A representative figure depicting the generic spectrum of the vector-like quarks in models with PC. For $Q=2/3$ and $-1/3$ the pair of dashed lines imply that either one of the lines is present in a particular model.}
	\label{fig:spectra}
\end{figure}

Singular value decomposition of $\mathcal{M}_{2/3}$ can be done numerically or perturbatively by expanding in powers of $\theta$.
The physically relevant cases yield masses for the top quark of either of the following types
\begin{align}
\label{top_mass}
m_t\propto \left\{ \begin{array}{ll}
\frac{f^2My_Ly_R\theta}{\sqrt{M^2+y_L^2f^2}\sqrt{M^2+y_R^2f^2}}\,, &\mbox{Type I}, \\ \noalign{\medskip} 
\frac{f^2y_Ly_R\theta}{\sqrt{M^2+y_{L,R}^2f^2}}\,, &\mbox{Type II}.
\end{array} \right.
\end{align}
In some cases, such as when both $|\omega^t_{L}(\theta)|$ and $|\omega^t_{R}(\theta)|\sim \mathcal{O}(\theta)$, the top quark mass is $\mathcal{O}(\theta^2)$ and must be discarded as it is too small.
In the left panel of Fig.\,\ref{fig:spectra} we show the contours satisfying $m_t=173$ GeV in the $y_R-y_L$ plane for different values of $M/f$. The solid (dashed) contours in the left panel of Fig.\,\ref{fig:spectra} represent models of Type I (II). Note that fairly large values for $y_L$ and $y_R$ are required to reproduce the correct top mass.

We now discuss the spectrum of the composite fermions with electric charge $Q=2/3$. The generic spectrum at tree level  is shown in the right panel of Fig.\,\ref{fig:spectra}. The presence of the $(n-1)\times(n-1)$ diagonal block ensures that $\mathcal{M}_{2/3}$ has $n-3$ exactly degenerate states with mass $M$ in its singular value decomposition.
This can be seen by noticing that the $\omega^t_{L,R}$ can be brought by a field redefinition to have only at most the first two components non-zero. 
The masses of the remaining two top-partners (recall that $n\times n$ mass matrix is composed of top quark and $n-1$ top-partners) are given by $\sqrt{M^2+y_{L}^2f^2}$ and $\sqrt{M^2+y_{R}^2f^2}$, for models of Type I. In case of Type II models one of the top-partners has a mass slightly heavier than $M$:  $\sqrt{M^2+y_{L,R}^2 v^2}$ (note that this leads to a very small mass splitting $\approx y^2_{L,R} v^2/2 M$), while the other is heavier with a mass of $\sqrt{M^2+y_{R,L}^2f^2}$.

\subsubsection*{Singular value decomposition of $\mathcal{M}_{-1/3}$:}

In the presence of a partner for the bottom quark leading to a non-zero  $\omega^b_R(\theta)$ the singular value decomposition of $\mathcal{M}_{-1/3}$ proceeds exactly as in the previous case. In the cases where there is no $b_R$ partner we assume that the mass of the bottom quark can be generated by some bilinear operator. Incorporating this assumption, we get a modified $\mathcal{M}_{-1/3}$ as
\begin{equation}
\label{m13n}
\mathcal{M}_{-1/3}=\left(\begin{array}{@{}c|c@{}}
\mu_b\theta
& \omega^b_L(\theta)^T \\
\hline
0_{n-1\times 1} &
M \mathbb{1}_{n-1}
\end{array}\right)\,,
\end{equation} 
where $\mu_b$ denotes the contribution from the bilinear operator of the type $\bar{q}_L\mathcal{O}b_R$. The generic expression for the bottom quark mass to $\mathcal{O}(\theta)$ can also be of two types as follows
\begin{align}
\label{bottom_mass}
m_b\propto \left\{ \begin{array}{lc}
\frac{\mu_b M\theta}{\sqrt{M^2+y_L^2f^2}}\,,  &\mbox{Type I}, \\ \noalign{\medskip} 
\mu_b\theta\,,  &\mbox{Type II}. 
\end{array} \right.
\end{align}
Eq.\,\eqref{m13n} now implies that there are $n-2$ degenerate states with mass equal to $M$. The remaining state can have a mass $\sqrt{M^2+y_L^2f^2}$ for Type I, and $\sqrt{M^2+y_L^2v^2}$ for Type II models.

\subsection{One loop self energy}

\begin{figure}[t]
	\centering
	\includegraphics[trim={2.5cm 24.5cm 1.8cm 1.5cm},clip,scale=1]{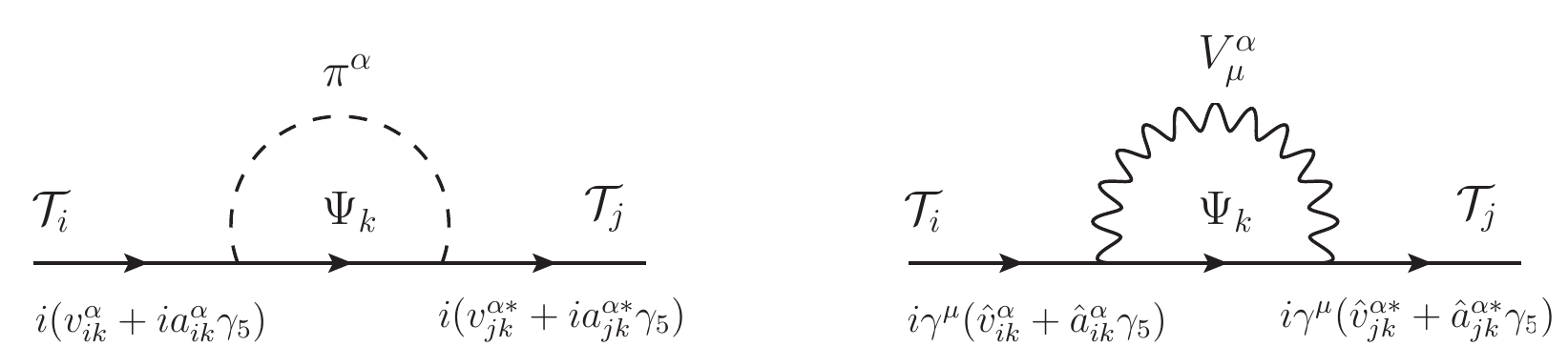}
	\caption{\small\it The one loop diagrams contributing to the self energy of the top-partners in terms of generic dimensionless couplings $v, a, \hat v, \hat a$ that can be expressed in terms of $y_L, y_R, \kappa, M$, and $f$ for a specific model. The expressions under the diagrams are the Feynman rules for the corresponding vertices.}
	\label{fig:self_energy}
\end{figure}
In this subsection we discuss the one loop self energy of the top-partners and its effect on their masses and widths.
Due to the presence of explicit breaking of $\mathcal{G}$ through gauge interactions and PC, we expect that the one loop corrections to the fermionic self energy lift the degeneracy in the top-partner spectrum. The relevant self energy diagrams are shown in Fig.\,\ref{fig:self_energy} 
where $\mathcal{T}_{i,j}$ represent the degenerate top-partners while $\Psi_k$ denotes either a partner or a SM quark running in the loop. The contributions from the two diagrams to the self-energy are
\begin{align}
 \label{self_energy_1}
 -i\Sigma_{ij}^\pi(\slashed p) &=\sum_{k,\alpha}\int_0^1 dx\int \frac{d^4l}{(2\pi)^4}\frac{\left[(\slashed px+M_k)v^\alpha_{ik}v^{\alpha*}_{jk}+(\slashed px-M_k)a^\alpha_{ik}a^{\alpha*}_{jk}+(\gamma_5-{\rm terms)}\right]}{(l^2-\Delta_\pi+i\epsilon)^2}\,,\\
 -i\Sigma_{ij}^V(\slashed p) &=\sum_{k,\alpha}\int_0^1 dx\int \frac{d^4l}{(2\pi)^4}\frac{2\left[(\slashed px-2M_k)\hat{v}^\alpha_{ik}\hat{v}^{\alpha*}_{jk}+(\slashed px+2M_k)\hat{a}^\alpha_{ik}\hat{a}^{\alpha*}_{jk}+(\gamma_5-{\rm terms)}\right]}{(l^2-\Delta_V+i\epsilon)^2}\,,
 \label{self_energy_2}
\end{align}
where $\Delta_{\pi,V}\equiv m_{\pi,V}^2x+(M_k^2-p^2x)(1-x)$. The $\gamma_5$ terms can be rotated away by a field redefinition and are ignored. The real part ${\rm Re}[\Sigma(\slashed p)]$ is logarithmically divergent and contributes to the mass correction. We regularize this divergence using a UV cut-off around the compositeness scale $\Lambda\sim4\pi f$, where the non-perturbative dynamics from the strongly interacting hypercolor sector kicks in. We denote the corrected central value of the mass of the semi-degenerate multiplet by $M_\mathcal{T}$ and the mass splitting by $\delta M_{ij}$ as 
\begin{equation}
M_\mathcal{T}=M+\frac{1}{N_\mathcal{T}}{\rm tr}({\rm Re}[\Sigma(M)])\,, \qquad  \delta M_{ij}={\rm Re}[\Sigma_{ij}(M)]-\frac{1}{N_\mathcal{T}}{\rm tr}({\rm Re}[\Sigma(M)])\delta_{ij}\,,
\end{equation}
where $N_\mathcal{T}$ denotes the number of degenerate states (typically $N_\mathcal{T}=2\mbox{ or }3$ in these models). 

Just as the mass splitting, the total decay width ($\Gamma_{\mathcal{T}}$) of these states can be expressed as a hermitian matrix with non-zero off-diagonal elements using the optical theorem as $\Gamma_{\mathcal{T}}=-2{\rm Im}[\Sigma]$. The $\Gamma_{\mathcal{T}}$ is the sum of the matrix valued partial decay widths for the appropriate channels as given below
\begin{align}
\Gamma_{ij}(\mathcal{T}\to\Psi_k\pi^\alpha) =v^\alpha_{ik}v^{\alpha*}_{jk}\Gamma^+_{\pi}+a^\alpha_{ik}a^{\alpha*}_{jk}\Gamma^-_{\pi}\,,\quad
\Gamma_{ij}(\mathcal{T}\to\Psi_k V^\alpha_\mu) =\hat{v}^\alpha_{ik}\hat{v}^{\alpha*}_{jk}\Gamma^-_{V}+\hat{a}^\alpha_{ik}\hat{a}^{\alpha*}_{jk}\Gamma^+_{V}\,,
\end{align}
where $\Gamma^\pm_{\pi}$ and $\Gamma^\pm_{V}$ are given as 
\begin{align}
\Gamma^{\pm}_{\pi} & =\frac{\lambda(M_\mathcal{T},M_k,m_\pi)}{16\pi M_\mathcal{T}}\left(1+\frac{M_k^2}{M_\mathcal{T}^2}-\frac{m_\pi^2}{M_\mathcal{T}^2} \pm 2\frac{M_k}{M_\mathcal{T}}\right),\\
\Gamma^{\pm}_{V} & =\frac{\lambda(M_\mathcal{T},M_k,M_V)M_\mathcal{T}}{16\pi M_V^2}\left[1+\frac{M_k^4}{M_\mathcal{T}^4}-2\frac{M_V^4}{M_\mathcal{T}^4}+\frac{ M_V^2}{M_\mathcal{T}^2}+\frac{M_k^2}{M_\mathcal{T}^2}\frac{M_V^2}{M_\mathcal{T}^2}-2\frac{M_k^2}{M_\mathcal{T}^2}\pm 6\frac{M_k}{M_\mathcal{T}}\frac{M_V^2}{M_\mathcal{T}^2}\right],
\end{align}
and, $\lambda(x,y,z)\equiv \sqrt{x^4+y^4+z^4-2x^2y^2-2y^2z^2-2x^2z^2}$ is the K\"all\'en function. 

Thus, the outcome of the analysis above is that in case of nearly degenerate states with $|\Gamma_{\mathcal{T}}|\sim |\delta M|\ll M_\mathcal{T}$ we obtain a matrix propagator
\begin{equation}
i\left[(\slashed p-M_\mathcal{T})\mathbb{1}+\frac{i}{2}(\Gamma_{\mathcal{T}}+2i\delta M)\right]_{ij}^{-1}
\approx\left[\frac{i(\slashed p + M_\mathcal{T})}{(p^2-M_\mathcal{T}^2)\mathbb{1}+i M_\mathcal{T}(\Gamma_{\mathcal{T}}+2i\delta M)}\right]_{ij}\equiv i(\slashed p + M_\mathcal{T})\Delta_{ij}(p^2).
\label{matrix_prop}
\end{equation}
When considering top-partners pair production and their subsequent decays one must take into account the interference between the channels involving the nearly degenerate states using the above matrix propagator, leading to an interesting computational challenge. We will show a way to handle this puzzle for the specific example discussed in the next section.

\section{Phenomenology of top-partners: an explicit example}
\label{pheno}

Having presented the general construction of these models in the previous sections we now move to describe a specific example of such models displaying various unusual phenomenological features. We choose to employ the $\rm SU(5)/SO(5)$ coset since it has a rich pNGB sector with anomalous couplings to dibosons leading to a wide variety of decay channels for the top-partners. 
This coset leads to 14 pNGBs whose decomposition  under $\rm SU(2)_L\times SU(2)_R \supset SU(2)_{cust}$ is given by\footnote{The superscripts denote the electric charges and $G$ denotes the Goldstones eaten by the SM gauge bosons.} (also see appendix\,\ref{sec_pngbs})
\begin{align}
{\mathbf {14}}&\to ({\mathbf 3},{\mathbf 3})+({\mathbf 2},{\mathbf 2})+({\mathbf 1},{\mathbf 1})\label{decofourteen}
\\&\to {\mathbf 1}(\chi_1^0) + {\mathbf 3}(\chi_3^{\pm},\chi_3^{0}) + {\mathbf 5}(\chi_5^{\pm\pm},\chi_5^{\pm},\chi_5^{0}) + {\mathbf 1}(h) +{\mathbf 3}(G^{\pm},G^{0}) + {\mathbf 1}(\eta). \nonumber
\end{align}
After chosing the coset, the remaining discrete choices to be made are the irreps for the fermionic partner and the spurions of the third quark family. 
We choose the following irreps for the construction of our model (see appendix~\ref{spurions} for details)
\begin{equation}
    \hat q_L = t_L D^1_{t_L} + b_L D^1_{b_L}  \in \mathbf{24},\quad  \hat t_R = t_R D^2_{t_R} \in \mathbf{24}, \quad  \Psi_A\in \mathbf{10}. \label{discretechoice}
\end{equation}
The spurions are involved in the construction of both the scalar potential and the Yukawa couplings, while the choice of the partner only affects the latter. The relevant parts of the Lagrangian are given in appendix~\ref{IR_realcase_theory}.

We choose $\Psi_A \in \mathbf{10}$ for simplicity, since it leads to a more restricted number of partners while retaining the main interesting features. Chosing the symmetric irrep for $\Psi$ would give rise to additional fermions with exotic charges $8/3$ and $-4/3$. 
There are only a few possible choices for the spurions~\cite{Agugliaro:2018vsu} satisfying the necessary requirements (such as no vevs for the triplets) and Eq.\,\eqref{discretechoice} complies with them.    

To set the notation we present the decomposition of the partners' irrep $\mathbf{10}_{2/3}$ under $\rm SU(2)_L\times SU(2)_R\times U(1)_X \supset SU(2)_L\times U(1)_Y$ 
\begin{align}
\mathbf{10}_{\frac{2}{3}} & \to (\mathbf{2},\mathbf{2})_{\frac{2}{3}}+ (\mathbf{3},\mathbf{1})_{\frac{2}{3}}+ (\mathbf{1},\mathbf{3})_{\frac{2}{3}}\,\label{decoten} \\
& \to  \mathbf{2}_{\frac{1}{6}} (T_{2\over 3},B_{-\frac{1}{3}})+\mathbf{2}_{\frac{7}{6}} (X_{5\over 3},X_{2\over 3})+\mathbf{3}_{\frac{2}{3}}(Y_{5\over 3},Y_{2\over 3},Y_{-\frac{1}{3}})+\mathbf{1}_{-\frac{1}{3}}(\tilde{B}_{-\frac{1}{3}})+\mathbf{1}_{\frac{2}{3}}(\tilde{T}_{2\over 3})+\mathbf{1}_{\frac{5}{3}}(\tilde{X}_{5\over 3})\,.\nonumber
\end{align}
The interesting novel feature of this model is having top partners with much reduced branching ratios to the usual SM channels $t\, h$, $t\, Z$, and $b\, W$ and instead a large branching ratio to beyond the SM (BSM) mediated channels such as $t (\eta \to \gamma\, \gamma)$ as we now proceed to discuss. 

\begin{figure}[t]
	\centering
	\includegraphics[width=0.6\textwidth,keepaspectratio]{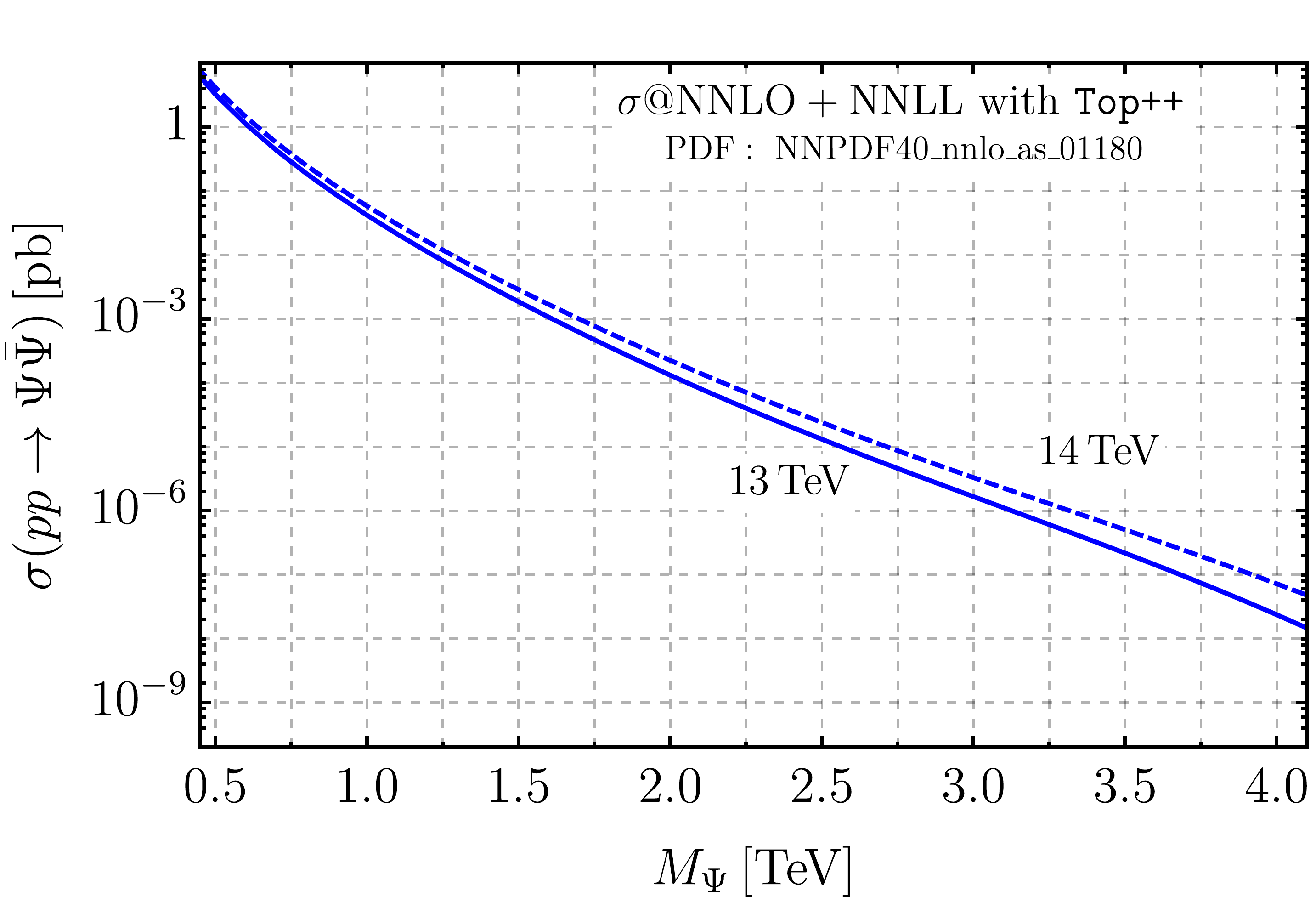}
	\caption{\small\it 
	The  pair production cross-section $\sigma(pp\to \Psi\bar{\Psi})$ for any color triplet calculated using \texttt{Top++}~\cite{Czakon:2011xx} at the NNLO+NNLL accuracy with centre of mass energies $\sqrt{s}=13$ TeV and $14$ TeV, respectively.}
	\label{fig:pair_prod}
\end{figure}

We consider only pair production of top-partners and their subsequent decays. 
All the possible two body decay channels for this model are listed in Table\,\ref{decay_channels}.
\begin{table}[t!]
	\begin{center}
\begin{tabular}{ccc}
\hline\hline
Top-partner & Decays to SM final states & Decays to BSM final states \\
\hline
$T_{\frac{2}{3}}, X_{\frac{2}{3}}, Y_{\frac{2}{3}}, \tilde T_{\frac{2}{3}}$	& $th$, $tZ$, $bW^+$	& $t \chi_{1,3,5}^0$, $t \eta$, $b \chi_{3,5}^+$   \\
$B_{-\frac{1}{3}}, Y_{-\frac{1}{3}}, \tilde B_{-\frac{1}{3}}$	& $tW^-$, $bh$, $bZ$	& $t \chi_{3,5}^-$, $b \chi_{1,3,5}^0$ , $b \eta$  \\
$X_{\frac{5}{3}}, Y_{\frac{5}{3}}, \tilde X_{\frac{5}{3}}$	& $tW^+$				& $t \chi_{3,5}^+$, $b \chi_5^{++}$ \\
\hline\hline
\end{tabular}
\caption{\small\it Possible decay channels of the vector-like fermionic partners for the model discussed in this section, with the choice given in Eq.\,\eqref{discretechoice}.}
\label{decay_channels}
\end{center}	
\end{table}
Recall that pair production of the vector-like quarks is model independent and is a function of their mass only. The  pair production cross-section $\sigma(pp\to \Psi\bar{\Psi})$ for any color triplet is calculated at the NNLO+NNLL accuracy with \texttt{Top++}~\cite{Czakon:2011xx} using the sets \texttt{NNPDF4.0} parton densities and is shown in Fig.\,\ref{fig:pair_prod} (see \cite{BuarqueFranzosi:2019dwg} for corrections arising from the finite size of the partners). 

We now need to consider the various decay modes. In this paper we are only concerned with the total cross section for each channel without any detailed study of the reach attainable at colliders. This is a first step needed to justify further studies and it already involves an interesting challenge, since we must deal with a nearly degenerate spectrum and want to take into account the interference between channels. This problem has been encountered and studied before in similar contexts~\cite{Cacciapaglia:2009ic,Barducci:2013zaa,Deandrea:2021vje}, but here we present a different angle on it, showing how to treat off-diagonal contributions to masses and widths in the particular limit of interest for this paper.

We quickly review the tree level spectrum of the partners (see appendix\,\ref{IR_realcase_theory} for explicit expressions of the mass matrix).
The three fermionic partners with $Q=5/3$ are all degenerate at tree level with mass $M$. 
In the $Q=-1/3$ sector we assume that the bottom quark receives a mass through a bilinear operator. The tree level masses of the remaining vector-like quarks with $Q=-1/3$ are $M$, $M$, and $\sqrt{M^2+y_L^2v^2}$, respectively.
The charge $2/3$ states in the gauge eigenbasis are denoted by $t, T_{\frac{2}{3}}, X_{\frac{2}{3}}, Y_{\frac{2}{3}}, \tilde T_{\frac{2}{3}}, $ where $t$ comes from the elementary sector and the remaining fields from Eq.\,\eqref{decoten}. The leading order mixing in the mass matrix appears from the coupling of $\tilde{T}_{\frac{2}{3}L}$ and $t_R$, while $Y_{\frac{2}{3}}$ mixes at $\mathcal{O}(\theta)$ and $T_{\frac{2}{3}}$, $X_{\frac{2}{3}}$ do not mix at all with either $t_L$ or $t_R$.  The tree level masses in the top sector after singular value decomposition are given as $m_{t},\, M,\, M,\, \sqrt{M^2+y_L^2v^2/4},$ and $\sqrt{M^2+y_R^2f^2}$, respectively. 

In what follows, we will be interested in the phenomenology of the lightest two top-partners $T_{\frac{2}{3}}$ and $X_{\frac{2}{3}}$ with tree level mass $M$. 
The benchmark parameters used for this study are displayed in the Table\,\ref{BP}. The pNGB masses are obtained by minimizing the potential given in Eq.\,\eqref{fullpot}. The benchmark shown in the Table\,\ref{BP} has mass eigenstates of the pNGBs approximately aligned to the custodial direction. Here $m_3$, $m_5$ and $m_1$ denote the masses of the custodial triplet $\chi_3$, quintet $\chi_5$ and the singlet $\chi_1^0$, while $m_\eta$ denotes the mass of the pure singlet $\eta$. 
\begin{table}[t!]
	\begin{center}
\begin{tabular}{cccccccccccc}
\hline\hline
\multicolumn{6}{c}{Masses (in GeV)} & & \multicolumn{3}{c}{Couplings}\\
\hline
 $f$ & $M$ & $m_3$ & $m_5$ & $m_1$& $m_\eta$ & & $y_L$ & $y_R$ & $\kappa$\\
\hline
$1000$ & $1500$ & $330$ & $315$ & $335$ & $290$ & & $1.80$ & $1.87$ & $0.50$ \\

\hline\hline
\end{tabular}
\caption{\small\it Choice of benchmark parameters used in this section.}
\label{BP}
\end{center}	
\end{table}

\subsection{Decay of nearly degenerate top-partners}
\label{tp_decay}

\begin{figure}[t]
	\centering
	\includegraphics[trim={5.5cm 18.5cm 5.5cm 2.8cm},clip,width=0.4\textwidth,keepaspectratio]{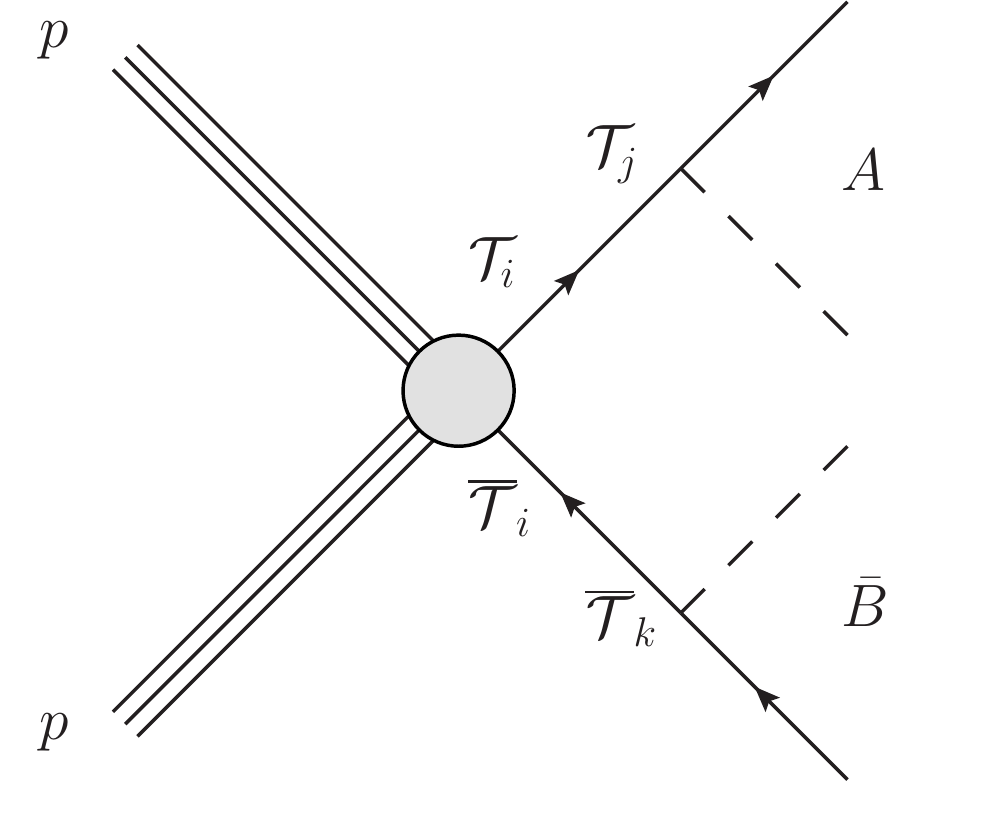}
	\caption{\small\it Feynman diagram for the process $pp \to \mathcal{T}\overline{\mathcal{T}} \to A\bar{B}$ where $\mathcal{T}$ denotes the lightest top-partners $T_{2/3}$ and $X_{2/3}$, while $A$ and $B$ represent any of the decay channels shown in Table\,\ref{decay_channels}. The propagators are labeled with e.g. $\mathcal{T}_i$, $\mathcal{T}_j$ to emphasize that the nearly degenerate top-partners can mix as shown in Eq.\,\eqref{matrix_prop}.}
	\label{fig:feymandiag}
\end{figure}
We consider the processes shown in the Fig.\,\ref{fig:feymandiag}  $pp\to \mathcal{T}\overline{\mathcal{T}}\to A \bar{B}$, where $\mathcal{T} \equiv T_{\frac{2}{3}}, X_{\frac{2}{3}}$ collectively denotes the two nearly degenerate top-partners and $A,B \equiv (bW^+),\, (tZ)\dots(b\chi_5^+)$  ($\bar B$ just being the charge conjugate of $B$). Thus $A$ and $B$  run over the possible SM and BSM decay modes shown in the first row of Table~\ref{decay_channels}. Since the pair production mode is universal we can factor out the production cross section 
$\sigma(pp\to T_{\frac{2}{3}}, \bar T_{-\frac{2}{3}}) = \sigma(pp\to X_{\frac{2}{3}}, \bar X_{-\frac{2}{3}})$ displayed in Fig.\,\ref{fig:pair_prod}. However, note that even at one-loop their mass splittings, computed using the Eqs.\,\eqref{self_energy_1} and \eqref{self_energy_2}, are smaller than their individual decay widths. Therefore we need to consider the interference between their identical decay channels and the fact that the propagator in Eq.\,\eqref{matrix_prop} is not diagonal. 

We have been able to show that in the narrow width approximation (NWA), even in the presence of an off-diagonal matrix propagator one can factorize the cross-section as 
\begin{align}
\sigma(pp\to\mathcal{T}\overline{\mathcal{T}}\to A\bar B)\overset{\mathrm{NWA}}{=}N_\mathcal{T}\sigma(pp\to \mathcal{T}\overline{\mathcal{T}}) \mathcal{BR}_2(\mathcal{T}\overline{\mathcal{T}}\to A \bar B)\, ,
\label{eq_BR2}
\end{align}
where in our case $N_\mathcal{T}\equiv 2$ denotes the number of degenerate fermionic states in the amplitude.
The quantity $\mathcal{BR}_2(\mathcal{T}\overline{\mathcal{T}}\to A \bar B)$ denotes the `joint branching ratio' of $\mathcal{T}\to A$ and $\overline{\mathcal{T}}\to \bar B$. 
This is the only factorization allowed in Eq.\,\eqref{eq_BR2} if one wants to keep the interference between the channels. The details of the calculation will be reported in an upcoming paper, however, the main steps for calculating $\mathcal{BR}_2(\mathcal{T}\overline{\mathcal{T}}\to A \bar B)$ are given in appendix\,\ref{decaywidths}.
Notice that it is not possible to write $\mathcal{BR}_2$ as a product of two branching ratios, as one would do if there was no degeneracy.
It is however possible to define an effective branching ratio of $\mathcal{T}\to A$ as
\begin{align}
\label{eff_BR}
\mathcal{BR}(\mathcal{T}\to A)\equiv \sum_{\bar B} \mathcal{BR}_2(\mathcal{T}\overline{\mathcal{T}}\to A \bar B)\,,~ 
{\rm such~that,}~ 
\sum_A \mathcal{BR}(\mathcal{T}\to A)=1\,, 
\end{align}
but again $\mathcal{BR}_2(\mathcal{T}\overline{\mathcal{T}}\to A \bar B)\ne \mathcal{BR}(\mathcal{T}\to A)\mathcal{BR}(\overline{\mathcal{T}}\to \bar B)$ implying non-trivial correlations between the two final states.

For our benchmark point the central value of the mass for $\mathcal{T} \equiv T_{\frac{2}{3}}, X_{\frac{2}{3}}$ including one loop corrections is found to be at $M_\mathcal{T}=1.36$ TeV. The numerical value of the one loop mass splitting and the total decay width matrix are
\begin{align}
\delta M=\left(\begin{array}{cc}
33.3 & -3.9 \\
-3.9 & -33.3
\end{array}\right) ~{\rm GeV},\quad
\Gamma_\mathcal{T}=\left(\begin{array}{cc}
121.8 & 3.7 \\
3.7 & 165.2
\end{array}\right) ~{\rm GeV}.
\end{align}

\begin{figure}[t]
	\centering
	\includegraphics[width=0.485\textwidth,keepaspectratio]{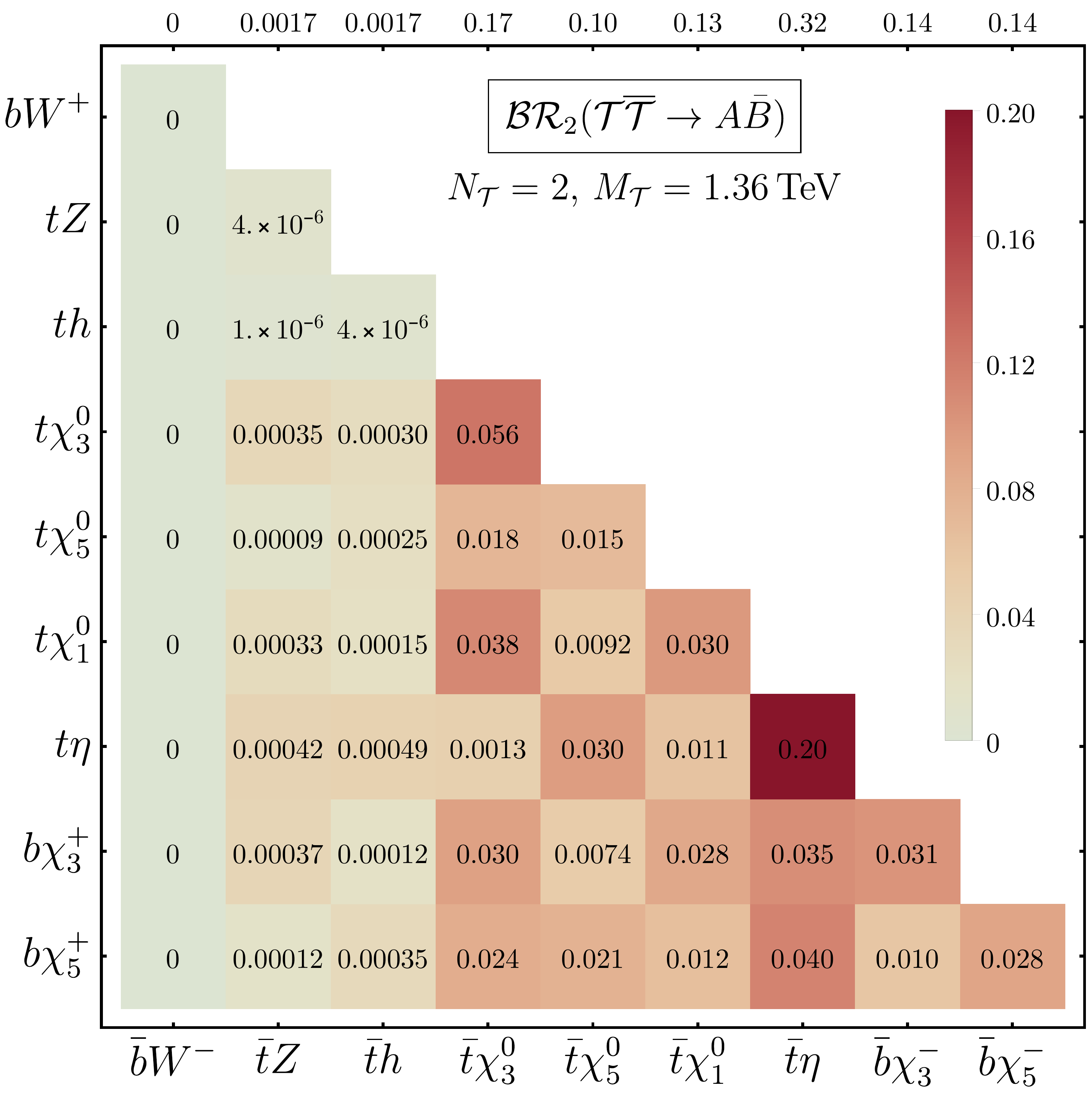}
	\caption{\small\it Joint branching ratios $\mathcal{BR}_2(\mathcal{T}\overline{\mathcal{T}}\to A \bar B)$ as in Eq.\,\eqref{eq_BR2} for our benchmark parameter choice where $A$ ($\bar{B}$) is given in the vertical (horizontal) axis of the plot. Since $\mathcal{BR}_2(\mathcal{T}\overline{\mathcal{T}}\to A \bar B)$ is a symmetric matrix we only present the lower triangle. The numbers on the top of the figure  show the values of $\mathcal{BR}(\mathcal{T}\to A)$ as given in Eq.\,\eqref{eff_BR}.}
	\label{fig:BR2}
\end{figure}

In Fig.\,\ref{fig:BR2} we present the $\mathcal{BR}_2(\mathcal{T}\overline{\mathcal{T}}\to A \bar B)$ for our choice of benchmark parameters. On top of the plot we also show the values of $\mathcal{BR}(\mathcal{T}\to A)$ as given in Eq.\,\eqref{eff_BR} for each two-body decay channels. The joint branching ratios are not too sensitive to the change in $y_L$, $\kappa$ and $M$ as long as $M\gg m_{\rm pNGB}$. For the model under consideration the couplings $\overline{\mathcal{T}}bW^+$ are zero at $\mathcal{O}(\theta)$ in the absence of PC mixing of $b_R$, and also $\overline{\mathcal{T}}tZ$ appears at $\mathcal{O}(\theta)$. From Fig.\,\ref{fig:BR2} we observe that the joint branching ratios to the BSM channels dominate compared to the SM channels. Clearly, the most promising channel is $pp\to \mathcal{T}\overline{\mathcal{T}}\to t\bar{t}\eta\eta$ with $\mathcal{BR}_2=0.20$. 

For the next-to-lightest top-partner $Y_{\frac{2}{3}}$ the dominating channels constitute of SM two-body final states. Similar calculations can be performed for the fermionic partners with $Q=-1/3$ and $Q=5/3$ as well. If the top-partners belong to $\mathbf{14}_{2/3}$ and the top quark is embedded in the adjoint of $\rm SU(5)$, a similar scenario can be constructed where the lightest top-partners also dominantly decay into the BSM states. 

\subsection{Decay of pNGBs}
\label{pngb_decay}

To extract the observable signal we need to combine $\sigma(pp \to \mathcal{T}\overline{\mathcal{T}}\to A \bar B)$ with the branching ratio of the pNGBs into the appropriate SM channels. While still promptly decaying, the pNGBs have a very small width so that in this case we can employ the usual NWA and ignore any further interference. The specific combination of cross-sections and branching ratios that needs to be employed depend on the final states one wants to target and how inclusive one wants to be.

As mentioned in the Introduction, we now focus on the diphoton channel and calculate the branching ratios of the neutral pseudoscalar pNGBs using the expressions given in \cite{Belyaev:2016ftv}. In the specific scenario we are considering, among the BSM pNGBs only $\eta$ couples to the $t\bar{t}$ pair with a strength (at $\mathcal{O}(\theta)$) proportional to 
\begin{equation}
\mathcal{L}_{\eta t\bar{t}}=\frac{f^2\,v\,\kappa\, y_L\, y_R^3}{2\sqrt{10}(M^2+y_R^2f^2)^{3/2}}i\eta \bar{t}\gamma_5t\,,
\end{equation}
while the other pseudoscalars $\chi_5^0$ and $\chi_1^0$ do not couple to $t\bar{t}$ at $\mathcal{O}(\theta)$. As a result, $\eta$ couples to a pair of gluons through a top quark loop which has a strength comparable to the WZW terms. 

\begin{figure}[t]
	\centering
	\includegraphics[width=0.485\textwidth,keepaspectratio]{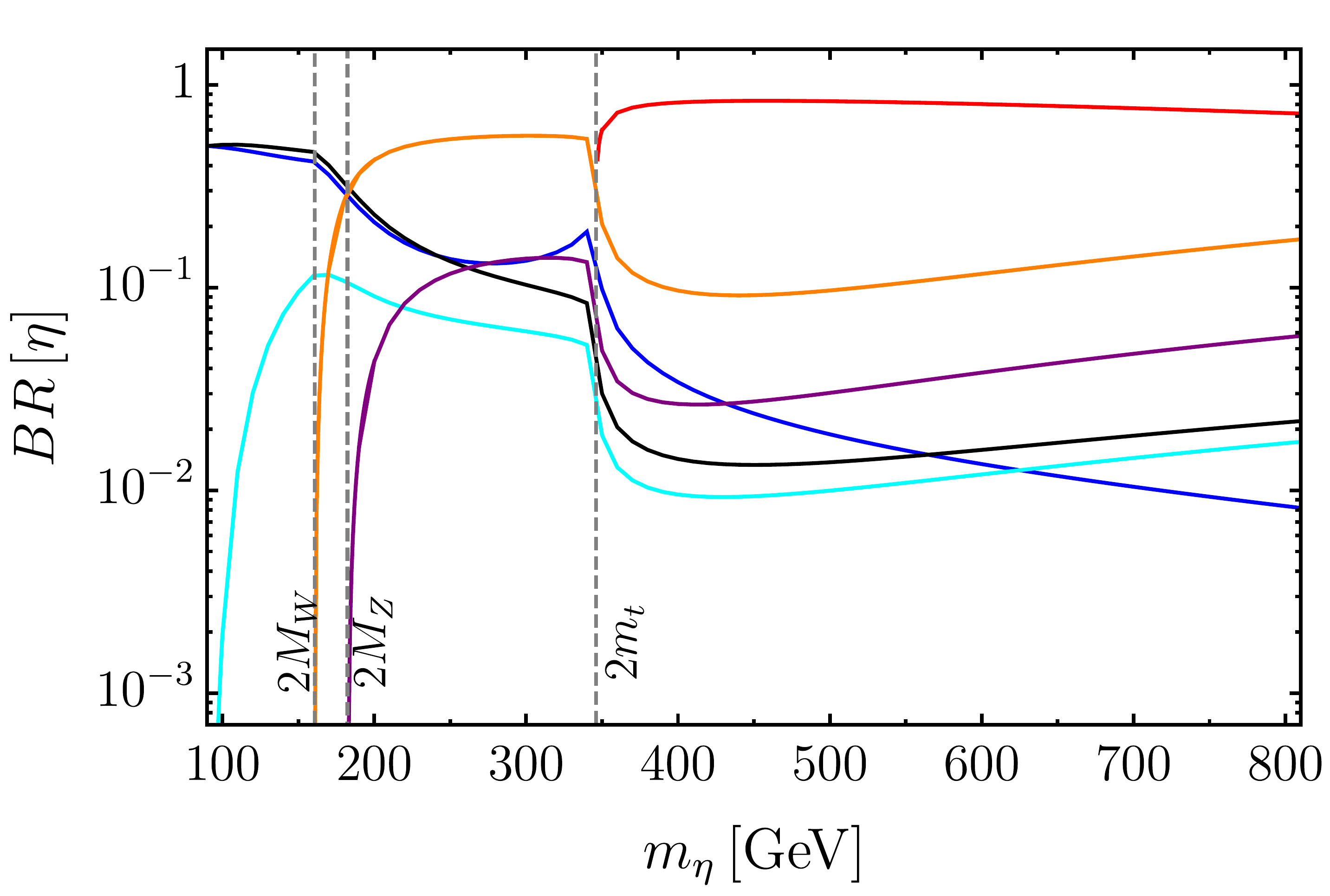}
	\includegraphics[width=0.485\textwidth,keepaspectratio]{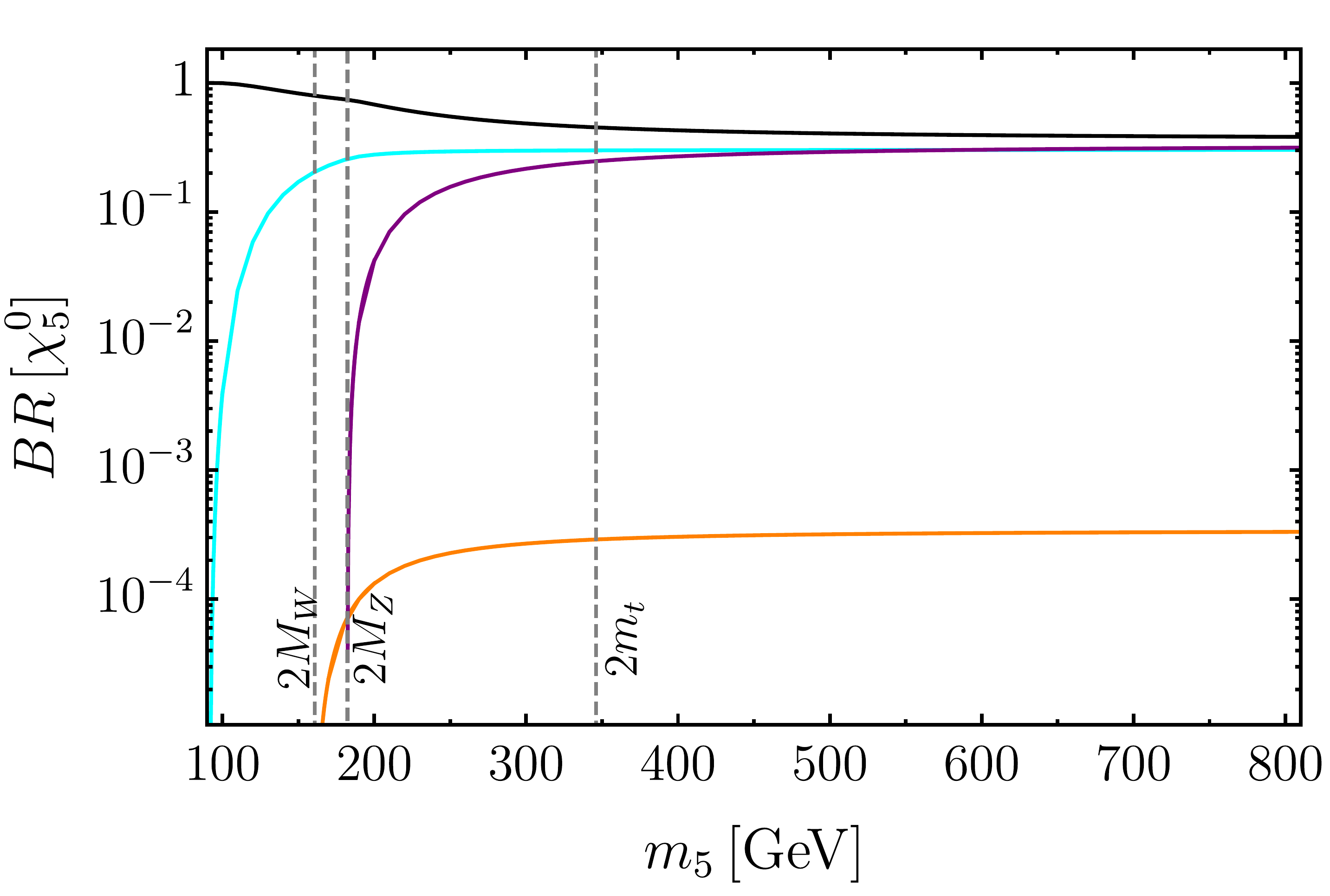}
	\includegraphics[width=0.605\textwidth,keepaspectratio]{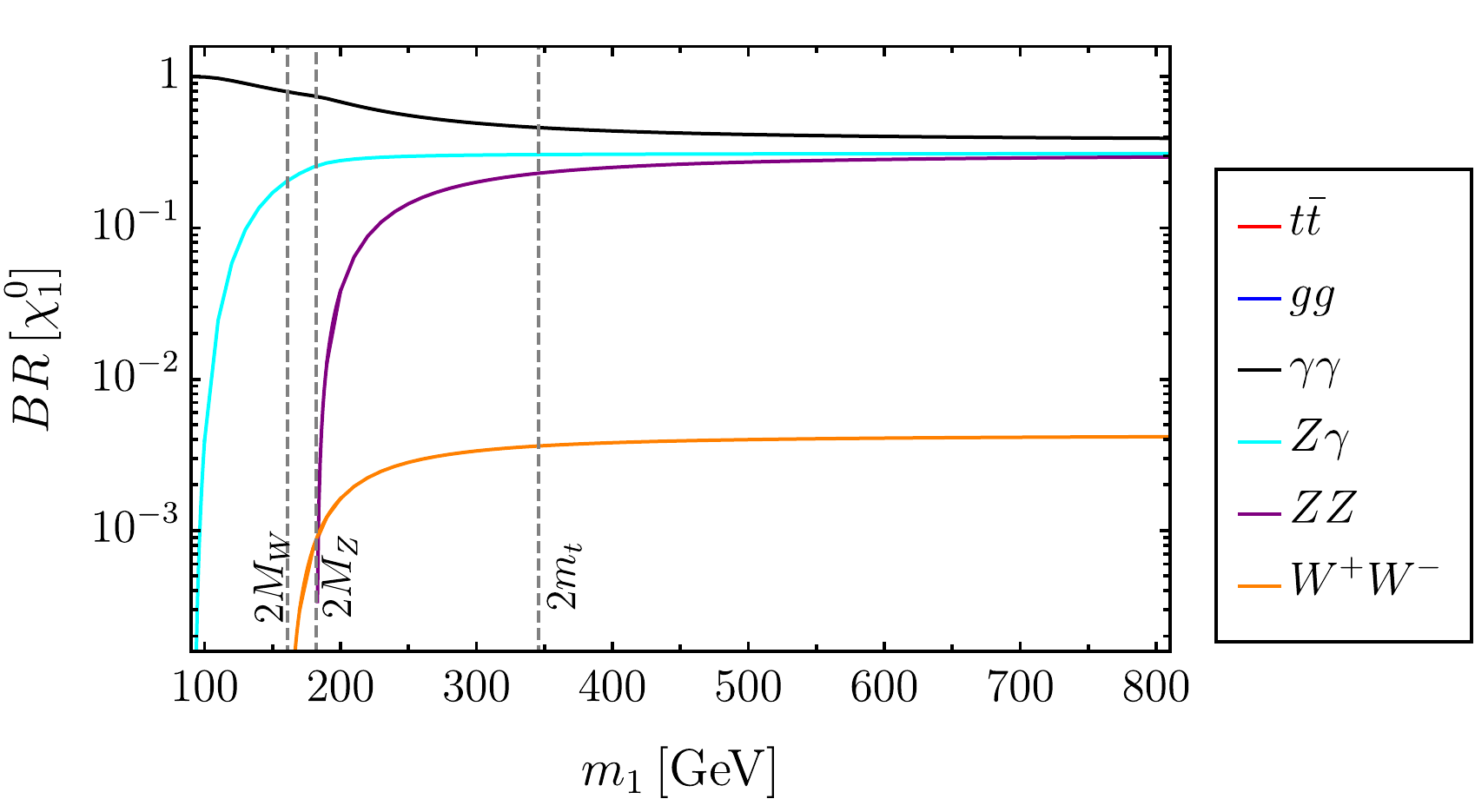}
	\caption{\small\it Branching ratios of pseudoscalar pNGBs into SM final states using the values of $M$, $f$, $y_{L,R}$ and $\kappa$ given in Table\,\ref{BP} and ${\rm dim}(\psi)=6$ in the anomalous couplings of Eq.\,\eqref{WZWterms}. We assume that mass splitting among the pNGBs remain same as given in  Table\,\ref{BP} throughout the entire mass range on the horizontal axes of each plots.}
	\label{fig:neutral_pngb_decay}
\end{figure}

In Fig.\,\ref{fig:neutral_pngb_decay} we present the branching ratios of $\eta$, $\chi^0_5$ and $\chi^0_1$, respectively using the benchmark values of $M$, $f$, $y_{L,R}$ and $\kappa$ given in Table\,\ref{BP}. Note that $\eta$ decays dominantly to $t\bar{t}$ pair if $m_\eta>2m_t$, while for $\chi_5^0$ and $\chi^0_1$ the branching ratio to diphoton dominates over a large range of masses. For our benchmark masses, the three body decays $\chi^0_1\to \chi_3^0 f\bar{f}$ and $\chi^0_1\to \chi^\pm_3f\bar{f}^\prime$ (where $f,f^\prime$ denote leptons or light quarks) via off-shell $W$ or $Z$ are orders of magnitude smaller compared to the two body decays ($BR(\chi^0_1\to \chi_3^0 f\bar{f})=3.3\times 10^{-4}$ and $BR(\chi^0_1\to \chi^\pm_3f\bar{f}^\prime)=1.2\times 10^{-4}$).   

Having the branching ratios of the neutral pNGBs we can compute the cross sections for various signals of interest. Focusing on the diphoton, we could consider the inclusive process with a top quark and a diphoton resonance in the final state:
\begin{align}
    \sigma\left(pp\to (t\gamma\gamma)\dots\right) &=\!\!\! \sum_{\pi^\alpha=\eta,\chi_{1,3,5}^0}\sum_{\bar{B}={\rm all}}\!\sigma\left(pp\to (t\pi^\alpha)\bar{B}\right)BR(\pi^\alpha\to\gamma\gamma), \nonumber\\
   & =N_\mathcal{T} \sigma(pp\to \mathcal{T}\overline{\mathcal{T}}) \!\!\!\!\sum_{\pi^\alpha=\eta,\chi_{1,3,5}^0}\!\!\mathcal{BR}\left(\mathcal{T}\to t\pi^\alpha\right)BR(\pi^\alpha\to\gamma\gamma),
   \label{xsectdiphot}
\end{align}
where, for ease of notation we now omit all the intermediate states in the expression of the cross section. For the specific scenario discussed above and for our choice of benchmark parameters we find $\sigma\left(pp\to (t\gamma\gamma)\dots\right)\sim 1.31$ fb. 

The cross section given in Eq.\,\eqref{xsectdiphot} is relevant for the leptonically decaying top quark where the distinction between a top and an anti-top is possible. However, for a hadronically decaying top, a more inclusive cross section as given below is of relevance:  
\begin{align}
    \sigma\left(pp\to (t/\bar{t}\,\gamma\gamma)\dots\right)  &=\!\!\!\! \sum_{\pi^\alpha=\eta,\chi_{1,3,5}^0}\!\left[\sum_{\bar{B}={\rm all}}\!\sigma\left(pp\to (t\pi^\alpha)\bar{B}\right) + \sum_{A={\rm all}}\!\sigma\left(pp\to A(\bar{t}\pi^\alpha)\right)\right] BR(\pi^\alpha\to\gamma\gamma),\nonumber\\ 
     &
     -\!\!\!\!\sum_{\pi^{\alpha,\beta}=\eta,\chi_{1,3,5}^0}\!\!\!\sigma\left(pp\to (t\pi^\alpha)(\bar{t}\pi^\beta)\right)BR(\pi^\alpha\to\gamma\gamma) BR(\pi^\beta\to\gamma\gamma), \nonumber\\
   & =N_\mathcal{T} \sigma(pp\to \mathcal{T}\overline{\mathcal{T}}) \left[ 2 \!\!\!\sum_{\pi^{\alpha}=\eta,\chi_{1,3,5}^0} \!\!\mathcal{BR}\left(\mathcal{T}\to t\pi^\alpha\right)BR(\pi^\alpha\to\gamma\gamma)\right., \nonumber\\ 
   & -\!\!\!\!\!\left.\sum_{\pi^{\alpha,\beta}=\eta,\chi_{1,3,5}^0}\!\!\!\mathcal{BR}_2\left(\mathcal{T}\overline{\mathcal{T}}\to (t\pi^\alpha)(\bar{t}\pi^\beta)\right)BR(\pi^\alpha\to\gamma\gamma) BR(\pi^\beta\to\gamma\gamma)\right].
\end{align}
We subtracted the processes with four photons in  the final state to avoid double counting. For our specific model, the choice given in Table\,\ref{BP} yields $\sigma\left(pp\to (t/\bar{t}\,\gamma\gamma)\dots\right)\sim 2.43$ fb. 

Note that more exclusive processes can be obtained by restricting the sum over $\bar B$ in Eq.\,\eqref{xsectdiphot}. On the other hand, more inclusive quantities, such as the fully inclusive diphoton signal $\sigma(pp\to \gamma\gamma\dots)$ would require the cross section of the processes involving additional intermediate fermionic partners with nearly equal masses, as well as the branching ratios of the charged pNGBs, which is beyond the scope of this paper.

\section{Conclusions}
\label{concl}

In this paper we constructed the low energy Lagrangian of a class of models with vector-like quarks and pNGB scalars, addressing the electroweak hierarchy problem within the partial compositeness framework. We presented the Lagrangian for three minimal cosets arising from strongly coupled gauge theories with fermionic matter in the UV. Our approach, based on symmetries motivated from specific UV realizations, greatly reduces the number of free parameters compared to the simplified models with the same field content. 

In these models, the structure of the fermion mass matrix and spectrum follows a specific pattern, and in particular, we highlighted the presence of nearly degenerate fermionic partners. We further estimated the one loop mass splitting in the degenerate sector.

We then focused on a concrete example based on the $\rm SU(5)/SO(5)$ coset to investigate the possible signatures of top-partners at colliders. We considered a minimalistic choice of both the irreps of the fermionic partners as well as the spurionic embedding of the SM third family quarks. We calculated the cross sections for production of a pair of lightest and nearly degenerate top-partners followed by their decays into SM and pNGB final states. The pNGBs were allowed to decay either into dibosons or a pair of third family quarks. For the nearly degenerate states we incorporated the effects of the off-diagonal self energy and the full quantum interference in the resonant production cross section and branching ratios. We showed that for the specific model in question the lightest top-partners decay dominantly into a top quark and a pseudoscalar pNGB, which in turn can decay to diphoton, primarily through anomalous interactions. This leads to a promising channel to search for top-partners at the LHC with an inclusive cross section of the order of a few femtobarns.   

\section*{Acknowledgments}

This work is supported by the Knut and Alice Wallenberg foundation under the grant KAW 2017.0100 (SHIFT project).
We would like to thank Luca Panizzi and Venugopal Ellajosyula for discussions and Thomas Flacke, Manuel Kunkel, and Leonard Schwarze for comments on the draft.

\appendix

\section{Building blocks for the IR theory}
\label{IR_theory}

In this section we provide more technical details on the basic building blocks needed to write the IR Lagrangian for a generic $\mathcal{G}/\mathcal{H}$ coset. The explicit expression for the generators of $\rm SU(2)_L\times SU(2)_R$ ($T_L^i, T_R^i$), the pNGB matrix $\Pi$, and the vacuum matrices $\Omega(\theta)$ are the same as in~\cite{Ferretti:2016upr} and will not be repeated.

\subsection{pNGBs}
\label{sec_pngbs}

We present the decompositions of the pNGBs under the various subgroups of $\mathcal{H}$, paying attention to the transformation to the custodial basis, to set the notation.

\subsubsection*{I. $\mathbf{\rm SU(4)/Sp(4)}$ coset:}

The minimal $\rm SU(4)/Sp(4)$ coset delivers 5 pNGBs \textit{viz.} a complex Higgs doublet $H=(H_+,H_0)$ and a real pseudoscalar singlet $\eta$, respectively. The decompositions of $\mathbf{5}$ of $\rm Sp(4)\simeq SO(5)$ under $\rm SU(2)_L\times SU(2)_R\supset SU(2)_L\times U(1)_Y$ is given as
\begin{align}
\label{decomp_5_14}
{\mathbf {5}} & \to ({\mathbf 2},{\mathbf 2})+({\mathbf 1},{\mathbf 1})\to {\mathbf 2}_{\pm 1/2}(H)+{\mathbf 1}_0(\eta)\,.
\end{align}
This is the same field content of the model~\cite{Gripaios:2009pe}.
In our convention the $\rm Sp(4)$ invariant tensor $\epsilon$ is the symplectic matrix $\epsilon_0\equiv i\sigma^3\otimes \sigma^2$ so that $U=\Sigma\epsilon_0\Sigma^T$. Because the additional pNGB $\eta$ is a singlet of $\rm SU(2)_L\times SU(2)_R$ nothing needs to be done to go to the custodial basis, other than, of course the usual decomposition already present in the SM $({\mathbf 2},{\mathbf 2})\to {\mathbf 1}(h) 
+ {\mathbf 3}(G^\pm, G^0)$, where $G^{\pm,0}$ are the would be Goldstone bosons eaten by the $W^\pm$ and $Z$.

\subsubsection*{II. $\mathbf{\rm SU(5)/SO(5)}$ coset:}

The 14 pNGBs in this coset transforming as $\mathbf{14}$ of $\rm SO(5)$ can be decomposed on restriction to $\rm SU(2)_L\times SU(2)_R\supset SU(2)_L\times U(1)_Y$ as
\begin{equation}
\label{decomp_14}
{\mathbf {14}}\to ({\mathbf 3},{\mathbf 3})+({\mathbf 2},{\mathbf 2})+({\mathbf 1},{\mathbf 1})\to {\mathbf 3}_0(\Phi_0)+{\mathbf 3}_{\pm 1}(\Phi_\pm)+{\mathbf 2}_{\pm 1/2}(H)+{\mathbf 1}_0(\eta)\,.
\end{equation}
Therefore the pNGBs comprise of the usual Higgs doublet $H=(H_+,H_0)$ along with a $\rm SU(2)_L$ real triplet $\Phi_0=(\phi_0^-,\phi_0^0,\phi_0^+)$, a complex triplet $\Phi_\pm=(\phi_\pm^-,\phi_\pm^0,\phi_\pm^+)$ and a real singlet $\eta$. 
For this coset the symmetric invariant matrix $\epsilon$ is taken to be the identity matrix so that $U=\Sigma\Sigma^T$. 

It is also relevant to show the pNGB content in the custodial basis, since this is a good approximate symmetry of the potental. Under ${\rm SU(2)_L\times SU(2)_R \to SU(2)_{cust}}$ the pNGBs decompose as
\begin{align}
\label{custodial_decomp}
({\mathbf 3},{\mathbf 3}) \to  {\mathbf 1}+{\mathbf 3}+{\mathbf 5}\,,\quad
({\mathbf 2},{\mathbf 2}) \to  {\mathbf 1}+{\mathbf 3}\,,\quad
({\mathbf 1},{\mathbf 1}) \to  {\mathbf 1}\,.
\end{align}
Since we assume that only the Higgs doublet receives a vev, the custodial triplet originating from the $({\mathbf 2},{\mathbf 2})$ serves as the would be Goldsotne boson to be eaten up by the $W^\pm$ and $Z$ bosons, as before. The physical pNGB spectrum contains one custodial quintet $\chi_5\equiv (\chi_5^0,\,\chi_5^\pm,\,\chi_5^{\pm\pm})$, one triplet $\chi_3\equiv(\chi_3^0,\,\chi_3^\pm)$, and three singlets ($h$, $\chi^0_1$, $\eta$), out of which the singlet $h$ associated with the $({\mathbf 2},{\mathbf 2})$ is identified with the 125 GeV Higgs boson.  The relation between the triplets $\Phi_0$ and $\Phi_\pm$ with the custodial basis ($\chi_1^0,\,\chi_3,\,\chi_5$) is given by
\begin{align}
\label{custodial_basis}
\nonumber
\phi_0^0 & =-\sqrt{\frac{2}{3}}\chi_5^0+\frac{1}{\sqrt{3}}\chi_1^0\,, & \phi_+^- & =\frac{1}{\sqrt{6}}\chi_5^0+\frac{1}{\sqrt{3}}\chi_1^0+\frac{i}{\sqrt{2}}\chi_3^0\,,\\
\phi_0^+ & =\frac{1}{\sqrt{2}}\left(\chi_5^++i\chi_3^+\right)\,, & 
\phi_+^0 & =\frac{1}{\sqrt{2}}\left(-\chi_5^++i\chi_3^+\right)\,, \qquad\qquad
\phi_+^+ =\chi^{++}_5\,.
\end{align}
Note that the phase convention for the Clebsch-Gordan coefficients in~\eqref{custodial_basis} is sligthly different from the conventional one for consistency with the choice of generators. The convention used here is the same as the one in~\cite{Agugliaro:2018vsu} apart from the sign of $\phi_0^+$.

\subsubsection*{III. $\mathbf{\complexcos}$ coset:}

Finally, in case of the $\mathbf{\complexcos}$ coset, discussed in~\cite{Ma:2015gra}, the 15 pNGBs transform under the adjoint of the unbroken $\rm SU(4)_d$ can be decomposed on restriction to $\rm SU(2)_L\times SU(2)_R\supset SU(2)_L\times U(1)_Y$ as
\begin{align}
\label{decomp_15_su4}
{\mathbf {15}} & \to ({\mathbf 3},{\mathbf 1})+({\mathbf 1},{\mathbf 3})+2\cdot({\mathbf 2},{\mathbf 2})+({\mathbf 1},{\mathbf 1})\to {\mathbf 3}_0(\Delta)+2\cdot{\mathbf 2}_{\pm 1/2}(H, H')+{\mathbf 1}_{\pm 1}(N^\pm)+2\cdot{\mathbf 1}_0(N^0, \eta)\,.
\end{align}
The $H$ is the same Higgs doublet as before, while $H'$ is an additional doublet. $\Delta=(\Delta^\pm, \Delta^0)$ is a triplet of $\rm SU(2)_L$ and $N=(N^\pm,N^0)$ a triplet of $\rm SU(2)_R$ decomposed according to its hypercharge, which in this case coincides with the electric charge. $\Delta$ and $N$ are already custodial triplets.

\subsection{Fermionic partners}
\label{top_partners}

\begin{table}[t!]
\def\arraystretch{1.7}
	\begin{center}
	\begin{footnotesize}
		\begin{tabular}{cllll}
		\hline\hline
			$\rm SO(5)\times U(1)_X$ &  $\rm SU(2)_L\times SU(2)_R \times U(1)_X$  &  $\rm SU(2)_L\times U(1)_Y$\\
			\hline
			
			$\mathbf{1}_{\frac{2}{3}}$ & $\to$ $
			(\mathbf{1},\mathbf{1})_{\frac{2}{3}}$ & $\to$ $\mathbf{1}_{\frac{2}{3}}$\\
			
			$\mathbf{5}_{\frac{2}{3}}$ & $\to$ $
			(\mathbf{1},\mathbf{1})_{\frac{2}{3}}+(\mathbf{2},\mathbf{2})_{\frac{2}{3}}$ & $\to$ $\mathbf{1}_{\frac{2}{3}}+\mathbf{2}_{\frac{1}{6}}+\mathbf{2}_{\frac{7}{6}}$\\
			
			$\mathbf{10}_{\frac{2}{3}}$ & $\to$ $(\mathbf{2},\mathbf{2})_{\frac{2}{3}}+ (\mathbf{3},\mathbf{1})_{\frac{2}{3}}+ (\mathbf{1},\mathbf{3})_{\frac{2}{3}}$ & $\to$ $ \mathbf{1}_{\frac{2}{3}}+\mathbf{1}_{\frac{5}{3}}+\mathbf{1}_{-\frac{1}{3}}+\mathbf{2}_{\frac{1}{6}}+\mathbf{2}_{\frac{7}{6}}+\mathbf{3}_{\frac{2}{3}}$\\
			
			$\mathbf{14}_{\frac{2}{3}}$ & $\to$ $(\mathbf{1},\mathbf{1})_{\frac{2}{3}}+ (\mathbf{2},\mathbf{2})_{\frac{2}{3}}+ (\mathbf{3},\mathbf{3})_{\frac{2}{3}}$ & $\to$ $\mathbf{1}_{\frac{2}{3}}+\mathbf{2}_{\frac{1}{6}}+\mathbf{2}_{\frac{7}{6}}+\mathbf{3}_{\frac{2}{3}}+\mathbf{3}_{\frac{5}{3}}+\mathbf{3}_{-\frac{1}{3}}$\\
			\hline 
			$\rm SU(4)\times U(1)_X$ & $\rm SU(2)_L\times SU(2)_R \times U(1)_X$  & $\rm SU(2)_L\times U(1)_Y$\\
			\hline
			
			$\mathbf{1}_{\frac{2}{3}}$ & $\to$ $(\mathbf{1},\mathbf{1})_{\frac{2}{3}}$ & $\to$ $\mathbf{1}_{\frac{2}{3}}$\\
			
			$\mathbf{6}_{\frac{2}{3}}$ & $\to$ $
			2\cdot(\mathbf{1},\mathbf{1})_{\frac{2}{3}}+(\mathbf{2},\mathbf{2})_{\frac{2}{3}}$ & $\to$ $2\cdot\mathbf{1}_{\frac{2}{3}}+\mathbf{2}_{\frac{1}{6}}+\mathbf{2}_{\frac{7}{6}}$\\
			
			$\mathbf{10}_{\frac{2}{3}}$ & $\to$ $(\mathbf{2},\mathbf{2})_{\frac{2}{3}}+ (\mathbf{3},\mathbf{1})_{\frac{2}{3}}+ (\mathbf{1},\mathbf{3})_{\frac{2}{3}}$ & $\to$ $ \mathbf{1}_{\frac{2}{3}}+\mathbf{1}_{\frac{5}{3}}+\mathbf{1}_{-\frac{1}{3}}+\mathbf{2}_{\frac{1}{6}}+\mathbf{2}_{\frac{7}{6}}+\mathbf{3}_{\frac{2}{3}}$\\
			
			$\mathbf{15}_{\frac{2}{3}}$ &  $\to$ $(\mathbf{1},\mathbf{1})_{\frac{2}{3}}+2\cdot(\mathbf{2},\mathbf{2})_{\frac{2}{3}}+ (\mathbf{3},\mathbf{1})_{\frac{2}{3}}+ (\mathbf{1},\mathbf{3})_{\frac{2}{3}}$ & $\to$ $ 2\cdot\mathbf{1}_{\frac{2}{3}}+\mathbf{1}_{\frac{5}{3}}+\mathbf{1}_{-\frac{1}{3}}+2\cdot\mathbf{2}_{\frac{1}{6}}+2\cdot\mathbf{2}_{\frac{7}{6}}+\mathbf{3}_{\frac{2}{3}}$\\
			\hline\hline
		\end{tabular}
		\end{footnotesize}
		\caption{\small\it The decomposition of $\rm Sp(4)\times U(1)_X\simeq SO(5)\times U(1)_X$ irreps and $\rm SU(4)\times U(1)_X$ irreps under $\rm SU(2)_L\times SU(2)_R\times U(1)_X$. All partners have $X=2/3$. In the last column $Y=T^3_R+X$. We do not include the $\mathbf{4}_{1/6}$ irreps since they present phenomenological problems~\cite{Agashe:2006at} and cannot be realized in the UV completions~\cite{Ferretti:2016upr,Belyaev:2016ftv} under consideration.}
		\label{tab:H_irreps}
	\end{center}	
\end{table}

The fermionic partners transform under an irrep of $\mathcal{H}$. In Table\,\ref{tab:H_irreps}, we present the decomposition of the relevant irreps of  $\rm SO(5)\times U(1)_X$ (same as those of $\rm Sp(4)\times U(1)_X$) and of $\rm SU(4)\times U(1)_X$ under $\rm SU(2)_L\times SU(2)_R\times U(1)_X \supset SU(2)_L\times U(1)_Y$. 
Below we show the explicit notations for the pseudo real, real and the complex cosets, respectively.

\subsubsection*{I. $\mathbf{\rm SU(4)/Sp(4)}$ coset:}

We consider partners in the $\mathbf{1}\, (\Psi_N)$, $\mathbf{5}\, (\Psi_A)$, $\mathbf{10}\, (\Psi_S/\Psi_D)$  of $\rm Sp(4)$. The explicit expressions for them are given in terms of $\rm SU(2)_L\times SU(2)_R$ submultiplets as (in $2+2$ block notation for the matrices)
\begin{align}
\label{top_partner_su4_sp4}
\Psi_N=\tilde{T}_{\frac{2}{3}}\,, 
\Psi_{A}=\left(\begin{array}{@{}c|c@{}}
\frac{i}{2}\tilde{T}_{\frac{2}{3}}\sigma^2
& \frac{1}{\sqrt{2}}\Psi_{(2,2)} \\
\hline
-\frac{1}{\sqrt{2}}\Psi_{(2,2)}^T &
\frac{i}{2}\tilde{T}_{\frac{2}{3}}\sigma^2
\end{array}\right), 
\Psi_{S}=\left(\begin{array}{@{}c|c@{}}
\Psi_{(3,1)}
& \frac{1}{\sqrt{2}}\Psi_{(2,2)} \\
\hline
\frac{1}{\sqrt{2}}\Psi_{(2,2)}^T &
\Psi_{(1,3)}
\end{array}\right),
\end{align}
where the submultiplets are given by
\begin{align}
\Psi_{(2,2)}=\left(\begin{array}{cc}
X_{\frac{5}{3}} & T_{\frac{2}{3}} \\
X_{\frac{2}{3}} & B_{-\frac{1}{3}}
\end{array}\right),
\Psi_{(3,1)}=\left(\begin{array}{cc}
Y_{\frac{5}{3}} & \frac{1}{\sqrt{2}}Y_{\frac{2}{3}} \\
\frac{1}{\sqrt{2}}Y_{\frac{2}{3}} & Y_{-\frac{1}{3}}
\end{array}\right),
\Psi_{(1,3)}=\left(\begin{array}{cc}
\tilde{X}_{\frac{5}{3}} & \frac{1}{\sqrt{2}}\tilde{T}_{\frac{2}{3}} \\
\frac{1}{\sqrt{2}}\tilde{T}_{\frac{2}{3}} & \tilde{B}_{-\frac{1}{3}}
\end{array}\right).
\end{align}
The adjoint irrep is equivalent to the symmetric, but the explicit form in terms of the fields is $\Psi_D=\Psi_S \epsilon_0$ since the transformations under a generic generator $T^a$ of $\rm Sp(4)$ are  respectively $T^a\Psi_S+\Psi_S T^{aT}$ and  ${[T^a, \Psi_D]}$, where $T^a\epsilon_0=-\epsilon_0T^{aT}$. 

\subsubsection*{II. $\mathbf{\rm SU(5)/SO(5)}$ coset:}

Similarly, for the real coset we consider partners in the $\mathbf{1}\, (\Psi_N)$, $\mathbf{5}\, (\Psi_F)$, $\mathbf{10}\, (\Psi_{A/D})$ and $\mathbf{14}\, (\Psi_S)$  of $\rm SO(5)$. The notations used for the partners belonging to irreps of $\rm SO(5)$ are (in $4+1$ block notation)
\begin{align}
\label{top_partner_su5_so5}
\nonumber
\Psi_N=\tilde{T}_{\frac{2}{3}}\,,\quad 
\Psi_F=\left(\begin{array}{c}\,
\Psi_{(2,2)}
\\
\hline
i\tilde{T}_{\frac{2}{3}}
\end{array}\right), \quad
\Psi_{A/D}=\left(\begin{array}{@{}c|c@{}}
\Psi_{(3,1)}+\Psi_{(1,3)}
& \frac{1}{\sqrt{2}}\Psi_{(2,2)} \\
\hline
-\frac{1}{\sqrt{2}}\Psi_{(2,2)}^T &
0
\end{array}\right), 
\end{align}
\begin{align}
\Psi_{S}=\left(\begin{array}{@{}c|c@{}}
\Psi_{(3,3)}+\frac{i}{2\sqrt{5}}\tilde{T}_{\frac{2}{3}}\mathbb{1}_4
& \frac{1}{\sqrt{2}}\Psi_{(2,2)} \\
\hline
\frac{1}{\sqrt{2}}\Psi_{(2,2)}^T &
-\frac{2i}{\sqrt{5}}\tilde{T}_{\frac{2}{3}}
\end{array}\right).
\end{align}
The $\rm SU(2)_L\times SU(2)_R$ submultiplets are
\begin{align}
\Psi_{(2,2)}=\frac{1}{\sqrt{2}}\left(
\begin{array}{c}
i B_{-\frac{1}{3}}-i X_{\frac{5}{3}} \\
B_{-\frac{1}{3}} + X_{\frac{5}{3}} \\
i T_{\frac{2}{3}} + i X_{\frac{2}{3}} \\
- T_{\frac{2}{3}} + X_{\frac{2}{3}} \\
\end{array}
\right),
\end{align}

\begin{align}
\Psi_{(3,1)}=\frac{1}{2}\left(\begin{array}{cccc}
0 & Y_{\frac{2}{3}} & \frac{Y_{-\frac{1}{3}}-Y_{\frac{5}{3}}}{\sqrt{2}} &  \frac{-iY_{-\frac{1}{3}}-iY_{\frac{5}{3}}}{\sqrt{2}}\\ 
& 0 & \frac{-iY_{-\frac{1}{3}}-iY_{\frac{5}{3}}}{\sqrt{2}} & \frac{-Y_{-\frac{1}{3}}+Y_{\frac{5}{3}}}{\sqrt{2}}  \\ 
&  & 0 & Y_{\frac{2}{3}}  \\ 
&  &  & 0 
\end{array}\right),
\Psi_{(1,3)}=\frac{1}{2}\left(\begin{array}{cccc}
0 & \tilde{T}_{\frac{2}{3}} & \frac{\tilde{B}_{-\frac{1}{3}}-\tilde{X}_{\frac{5}{3}}}{\sqrt{2}} &  \frac{i\tilde{B}_{-\frac{1}{3}}+i\tilde{X}_{\frac{5}{3}}}{\sqrt{2}}\\ 
& 0 & \frac{-i\tilde{B}_{-\frac{1}{3}}-i\tilde{X}_{\frac{5}{3}}}{\sqrt{2}} & \frac{\tilde{B}_{-\frac{1}{3}}-\tilde{X}_{\frac{5}{3}}}{\sqrt{2}}  \\ 
&  & 0 & -\tilde{T}_{\frac{2}{3}}  \\ 
&  &  & 0 
\end{array}\right),
\label{psi_31}
\end{align}

\begin{align}
\Psi_{(3,3)}=\frac{1}{2}\left(\begin{array}{cccc}
iY_{\frac{2}{3}}+iV_{-\frac{4}{3}}-U_{\frac{8}{3}} & V_{-\frac{4}{3}}-iU_{\frac{8}{3}} & \frac{V_{-\frac{1}{3}}-U_{\frac{5}{3}}-Y_{-\frac{1}{3}}+Y_{\frac{5}{3}}}{\sqrt{2}} &  \frac{iV_{-\frac{1}{3}}+iU_{\frac{5}{3}}+iY_{-\frac{1}{3}}+iY_{\frac{5}{3}}}{\sqrt{2}}\\ 
& iY_{\frac{2}{3}}-iV_{-\frac{4}{3}}+U_{\frac{8}{3}} & \frac{-iV_{-\frac{1}{3}}-iU_{\frac{5}{3}}+iY_{-\frac{1}{3}}+iY_{\frac{5}{3}}}{\sqrt{2}} & \frac{V_{-\frac{1}{3}}-U_{\frac{5}{3}}+Y_{-\frac{1}{3}}-Y_{\frac{5}{3}}}{\sqrt{2}}\\
&  & -iY_{\frac{2}{3}}+iU_{\frac{2}{3}}+iV_{\frac{2}{3}} & U_{\frac{2}{3}}-V_{\frac{2}{3}}  \\ 
&  &  & -iY_{\frac{2}{3}}-iU_{\frac{2}{3}}-iV_{\frac{2}{3}}
\end{array}\right).
\label{psi_33}
\end{align}
We present only the upper triangular matrices in Eqs.\,\eqref{psi_31} and \eqref{psi_33}. One should remember that $\Psi_{(3,1)}$ and $\Psi_{(1,3)}$ are antisymmetric while $\Psi_{(3,3)}$ is symmetric. For this coset the notation for the adjoint and antisymmetric fields are exactly the same since we choose $T^a=-T^{aT}$ for all $\rm SO(5)$ generators. 

\subsubsection*{III. $\mathbf{\rm SU(4)_l\times SU(4)_r/SU(4)_d}$ coset:} 
The field content of the complex coset can be inferred directly from that of the pseudoreal one, since the generators of $\rm SU(2)_L\times SU(2)_R$ are the same. Specifically, the antisymmetric of ${\rm SU(4)_d}$ is the same as that of ${\rm Sp(4)}$ in \eqref{top_partner_su4_sp4} augmented by an additional $\rm Sp(4)$ singlet $\frac{1}{2}T'_{\frac{2}{3}}\epsilon_0$, the symmetric is exactly the same, and the adjoint can be taken as $\Psi_A\epsilon_0 + \Psi_S\epsilon_0$ for two generic symmetric and antisymmetric matrices as in \eqref{top_partner_su4_sp4}. 

\subsection{Spurions}
\label{spurions}
Finally we list the explicit expressions for the spurion matrices realizing the embedding of the elementary fields $t_L$, $b_L$ and $t_R$. The ones relevant for section~\ref{pheno} are the adjoint spurions $D^1_{t_L}$, $D^1_{b_L}$, and $D^2_{t_R}$ of $\rm SU(5)/SO(5)$ coset as discussed below. 
Notice that all spurions must carry $X=2/3$ since we need to form a Yukawa coupling with the fermionic partners by PC. This restricts the possible embeddings quite significantly. In particular, looking at the decomposition of the various irreps under $\rm SU(2)_L\times SU(2)_R$,  the only allowed embedding for the LH doublet $q_L$ is in the $(\mathbf{2},\mathbf{2})_{\frac{2}{3}}$, while $t_R$ can be embedded into  the singlet $(\mathbf{1},\mathbf{1})_{\frac{2}{3}}$ or the $T^3_R=0$ component of $(\mathbf{1},\mathbf{3})_{\frac{2}{3}}$.

\subsubsection*{I. $\mathbf{\rm SU(4)/Sp(4)}$ coset:}

The relevant spurions are those included in the irreps of $\rm SU(4)$: ${\mathbf{1}},{\mathbf{6}},{\mathbf{10}},\bar{\mathbf{10}},{\mathbf{15}}$, 
where the symmetric cases ${\mathbf{10}},\bar{\mathbf{10}}$ can be treated together. 
The embedding matrices for $t_R$ are given by (with the superscripts denoting possible degeneracy in the embedding)
\begin{align}
\nonumber
&N_{t_R}=(1), ~~
A_{t_R}^1 = \frac{i}{2}\left( \begin{matrix}\sigma^2 & 0\\0 & -\sigma^2 \end{matrix}\right), ~~
A_{t_R}^2 = \frac{i}{2}\left( \begin{matrix}\sigma^2 & 0\\0 & \sigma^2 \end{matrix}\right), ~~
S_{t_R} = \frac{1}{\sqrt{2}}\left( \begin{matrix} 0 & 0\\0 & \sigma^1 \end{matrix}\right), 
\end{align}
\begin{align}
D_{t_R}^1 = \frac{1}{2}\left( \begin{matrix}{\mathbb{1}}_2 & 0\\0 & -{\mathbb{1}}_2  \end{matrix}\right),~~
D_{t_R}^2 = \frac{1}{\sqrt{2}}\left( \begin{matrix}0 & 0\\0 & \sigma^3 \end{matrix}\right).
\label{spurion_tR_SU4}
\end{align}
The antisymmetric matrices $A_{t_R}^{1,2}$ and the adjoint $D^1_{t_R}$ are in the $(\mathbf{1},\mathbf{1})$. ($A_{t_R}^1$ is a full $\rm Sp(4)$ singlet proportional to the invariant tensor $\epsilon_0$.) The symmetric matrix  $S_{t_R}$ and the adjoint $D^2_{t_R}$ are $T_R^3=0$ components of $(\mathbf{1},\mathbf{3})$.
The matrices for embedding $t_L$ and $b_L$ are
\begin{footnotesize}
\begin{align}
\label{spurion_qL_SU4}
 A_{t_L} \!= \frac{1}{\sqrt{2}}\!\left( \begin{matrix}0 & 0 & 0 & 1\\0 & 0 & 0 & 0\\0 & 0 & 0 & 0\\ -1 & 0 & 0 & 0\end{matrix}\right), A_{b_L} \!= \frac{1}{\sqrt{2}}\!\left( \begin{matrix}0 & 0 & 0 & 0\\0 & 0 & 0 & 1\\0 & 0 & 0 & 0\\ 0 & -1 & 0 & 0\end{matrix}\right),
S_{t_L} \!= \frac{1}{\sqrt{2}}\!\left( \begin{matrix}0 & 0 & 0 & 1\\0 & 0 & 0 & 0\\0 & 0 & 0 & 0\\ 1 & 0 & 0 & 0\end{matrix}\right), S_{b_L} \!= \frac{1}{\sqrt{2}}\!\left( \begin{matrix}0 & 0 & 0 & 0\\0 & 0 & 0 & 1\\0 & 0 & 0 & 0\\ 0 & 1 & 0 & 0\end{matrix}\right), 
\end{align}
\end{footnotesize}
and for the adjoint $D^1_{t_L}=S_{t_L}\epsilon_0$,  $D^2_{t_L}=A_{t_L}\epsilon_0$, $D^1_{b_L} = S_{b_L}\epsilon_0$, $D^2_{b_L} = A_{b_L}\epsilon_0$.

\subsubsection*{II. $\mathbf{\rm SU(5)/SO(5)}$ coset:}

Similarly, for the real case, we need to consider spurions in the ${\mathbf{1}},{\mathbf{5}},{\mathbf{10}},{\mathbf{15}},{\mathbf{24}}$ of $\rm SU(5)$ where, again, the conjugate irreps give rise to the same spurions.
With the same notation as in the previous case, we have, for $t_R$
\begin{footnotesize}
\begin{align}
N_{t_R}\!=(1),~
F_{t_R}\!=\left(\!\begin{array}{cc}
0 \\
0\\
0\\
0\\
1
\end{array}\!\right),~
A_{t_R} \!=\frac{i}{2}\left(\!\begin{array}{@{}c|c@{}}
\begin{matrix}
\sigma^2 & 0_{2\times 2} \\
0_{2\times 2} & -\sigma^2
\end{matrix}
& 0_{4\times 1} \\
\hline
0_{1\times 4} &
0
\end{array}\!\right),~
S^1_{t_R}\!=\frac{1}{2\sqrt{5}}\!\left(\!\begin{array}{@{}c|c@{}}
\begin{matrix}
\mathbb{1}_2 & 0_{2\times 2} \\
0_{2\times 2} & \mathbb{1}_2
\end{matrix}
& 0_{4\times 1} \\
\hline
0_{1\times 4} &
-4
\end{array}\!\right),~
S^2_{t_R}\!=\frac{1}{\sqrt{5}}\mathbb{1}_5.
\end{align}
\end{footnotesize}
The matrices in the adjoint are given by $D^1_{t_R}=S^1_{t_R}$ and $D^2_{t_R}=A_{t_R}$.
As far as the spurions for the LH fields are concerned we use
\begin{footnotesize}
\begin{align}
\nonumber
F_{b_L}=\frac{1}{\sqrt{2}}\left(\begin{matrix}
i \\
1 \\
0 \\
0\\
0
\end{matrix}\right)\,, \qquad\qquad
F_{t_L}=\frac{1}{\sqrt{2}}\left(\begin{matrix}
0 \\
0 \\
i \\
-1\\
0
\end{matrix}\right)\,,
\end{align}
\begin{align}
A_{b_L}=\frac{1}{2}\left(\begin{array}{@{}c|c@{}}
0_{4\times 4} & \begin{matrix}
i \\
1 \\
0 \\
0
\end{matrix} \\
\hline
\begin{matrix}
-i & -1 & 0 & 0
\end{matrix} 
& 0
\end{array}\right)\,, \quad
A_{t_L}=\frac{1}{2}\left(\begin{array}{@{}c|c@{}}
0_{4\times 4} & \begin{matrix}
0 \\
0 \\
i \\
-1
\end{matrix} \\
\hline
\begin{matrix}
0 & 0 & -i & 1
\end{matrix} 
& 0
\end{array}\right)\,,
\nonumber
\end{align}
\end{footnotesize}
\begin{footnotesize}
\begin{align}
S_{b_L}=\frac{1}{2}\left(\begin{array}{@{}c|c@{}}
0_{4\times 4} & \begin{matrix}
i \\
1 \\
0 \\
0
\end{matrix} \\
\hline
\begin{matrix}
i & 1 & 0 & 0
\end{matrix} 
& 0
\end{array}\right)\,, \quad
S_{t_L}=\frac{1}{2}\left(\begin{array}{@{}c|c@{}}
0_{4\times 4} & \begin{matrix}
0 \\
0 \\
i \\
-1
\end{matrix} \\
\hline
\begin{matrix}
0 & 0 & i & -1
\end{matrix} 
& 0
\end{array}\right)\,.
\end{align}
\end{footnotesize}
with the matrices in the adjoint given by $D^1_{t_L}=S_{t_L}$, $D^2_{t_L}=A_{t_L}$, $D^1_{b_L}=S_{b_L}$, $D^2_{b_L}=A_{b_L}$.

\subsubsection*{III. $\mathbf{\rm SU(4)_l\times SU(4)_r/SU(4)_d}$ coset:}

Lastly we consider the complex case. Here, in principle, we need to specify the irrep of both $\rm SU(4)_{l,r}$ but, again, the explicit matrices are independent on these details and we can lump, e.g. $({\mathbf{1}},{\mathbf{10}}), ({\mathbf{1}},\bar{\mathbf{10}}), ({\mathbf{10}},{\mathbf{1}}), (\bar{\mathbf{10}},{\mathbf{1}})$ into a $S$ of $\rm SU(4)$. Having reduced the problem to a single $\rm SU(4)$ the spurions $N,A,S$, and $D$ in the complex case are given by the same numerical expressions as in the pseudoreal case, \eqref{spurion_tR_SU4} and \eqref{spurion_qL_SU4}, and need not be repeated.
In addition, we must also consider the $B$ifundamental matrices $B$. 
$({\mathbf{4}},{\mathbf{4}}), (\bar{\mathbf{4}},\bar{\mathbf{4}})$ are given by the same expression as the symmetric and the antisymmetric given in the pseudoreal case. Similarly, $({\mathbf{4}},\bar {\mathbf{4}}), (\bar{\mathbf{4}},{\mathbf{4}})$ are the same as the adjoint and an additional singlet only viable for the RH top ($B_{t_R}=\frac{1}{2}{\mathbf{1}}_4$).

\section{The Lagrangian of the model in section\,\ref{pheno}}
\label{IR_realcase_theory}

The interaction Lagrangian for the model presented in section\,\ref{pheno} involving pNGB scalars in the custodial basis and the fermions in the gauge basis can be written as  
\begin{align}
\nonumber
\mathcal{L}_{\mathcal{T}-\pi} &= \overline{\mathcal{T}}_{L}\left[-\mathcal{M}_{2/3}+\sum_{i}\pi^0_i\mathcal{I}_{\pi^0_i}\right]\mathcal{T}_{R}+{\rm h.c.}+\sum_{j}\pi^+_j\left[\overline{\mathcal{T}}_{L}\mathcal{I}^1_{\pi^+_j}\mathcal{B}_{R}+\overline{\mathcal{T}}_{R}\mathcal{I}^2_{\pi^+_j}\mathcal{B}_{L}\right]+{\rm h.c.}\\
\nonumber
& +\sum_{j}\pi^+_j\left[\overline{\mathcal{Y}}_{R}\hat{\mathcal{I}}^1_{\pi^+_j}\mathcal{T}_{L}+\overline{\mathcal{Y}}_{L}\hat{\mathcal{I}}^2_{\pi^+_j}\mathcal{T}_{R}\right]+{\rm h.c.}+\sum_{i}\partial_\mu \pi^0_i\overline{\mathcal{T}}\gamma^\mu\mathcal{I}_{\partial \pi^0_i}\mathcal{T} \\ 
& +\sum_{j}\partial_\mu\pi^+_j\left[\overline{\mathcal{T}}\gamma^\mu\mathcal{I}_{\partial\pi^+_j}\mathcal{B}+\overline{\mathcal{Y}}\gamma^\mu\hat{\mathcal{I}}_{\partial\pi^+_j}\mathcal{T}\right]+{\rm h.c.}
\label{L_int}
\end{align}
where $\pi_i^0\equiv h,\chi^0_3,\chi^0_5,\chi^0_1,\eta$ and $\pi^+_j\equiv \chi^+_3,\chi^+_5$ and
\begin{align}
\mathcal{T}\equiv\left(t, T_{\frac{2}{3}}, X_{\frac{2}{3}}, \tilde{T}_{\frac{2}{3}}, Y_{\frac{2}{3}}\right)^T, \quad \mathcal{B}\equiv\left(b, B_{-\frac{1}{3}}, Y_{-\frac{1}{3}}, \tilde{B}_{-\frac{1}{3}}\right)^T, \quad \mathcal{Y}\equiv\left(X_{\frac{5}{3}}, Y_{\frac{5}{3}}, \tilde{X}_{\frac{5}{3}}\right)^T. 
\end{align} 
In contrast to the section\,\ref{pheno}, in this appendix we denote the column vector containing all fermions with $Q=2/3$ as $\mathcal{T}$. The interactions between the weak gauge bosons and the fermionic partners are similarly given by the follwoing Lagrangian.
\begin{align}
\mathcal{L}_{\mathcal{T}-W/Z} &= \frac{e}{s_Wc_W}\bigg[\overline{\mathcal{T}}_{L}\gamma^\mu\mathcal{I}^L_{Z}\mathcal{T}_{L}+\overline{\mathcal{T}}_{R}\gamma^\mu\mathcal{I}^R_{Z}\mathcal{T}_{R}\bigg]Z_\mu+\frac{e}{s_W}\bigg[\overline{\mathcal{T}}_{L}\gamma^\mu\mathcal{I}^L_{W}\mathcal{B}_{L}+\overline{\mathcal{T}}_{R}\gamma^\mu\mathcal{I}^R_{W}\mathcal{B}_{R}\bigg]W^+_\mu+{\rm h.c.} \nonumber\\
& +\frac{e}{s_W}\bigg[\overline{\mathcal{Y}}_{L}\gamma^\mu\hat{\mathcal{I}}^L_{W}\mathcal{T}_{L}+\overline{\mathcal{Y}}_{R}\gamma^\mu\hat{\mathcal{I}}^R_{W}\mathcal{T}_{R}\bigg]W^+_\mu+{\rm h.c.}
\label{L_int_gauge}
\end{align}
\begin{table}[t!]
\begin{center}
\begin{footnotesize}
    \centering
    \begin{tabular}{lllllllll}
    \hline\hline
        $\pi^0_i$ & & \thead{$\mathcal{I}_{\pi^0_i}$} & & \thead{$\mathcal{I}_{\partial\pi^0_i}$} \\
        \hline
        $h$ & & $\left(
\begin{array}{ccccc}
 0 & 0 & 0 & \frac{y_L c_\theta}{2}   & -\frac{y_L c_\theta}{2}  \\
 0 & 0 & 0 & 0 & 0 \\
 0 & 0 & 0 & 0 & 0 \\
 -\frac{y_R s_\theta}{2}  & 0 & 0 & 0 & 0 \\
 \frac{y_R s_\theta }{2} & 0 & 0 & 0 & 0 \\
\end{array}
\right)$ & & $\left(
\begin{array}{ccccc}
 0 & 0 & 0 & 0 & 0 \\
 0 & 0 & 0 & -\frac{i \kappa}{4 f} & \frac{i \kappa}{4 f} \\
 0 & 0 & 0 & -\frac{i \kappa}{4 f} & \frac{i \kappa}{4 f} \\
 0 & \frac{i \kappa}{4 f} & \frac{i \kappa}{4 f} & 0 & 0 \\
 0 & -\frac{i \kappa}{4 f} & -\frac{i \kappa}{4 f} & 0 & 0 \\
\end{array}
\right)$\\ 

$\chi^0_3$ & & $\left(
\begin{array}{ccccc}
 0 & y_L s^2_{\theta/2} & y_L c^2_{\theta/2} & 0 & 0 \\
 \frac{y_R s_\theta}{2}   & 0 & 0 & 0 & 0 \\
 -\frac{y_R s_\theta}{2}  & 0 & 0 & 0 & 0 \\
 0 & 0 & 0 & 0 & 0 \\
 0 & 0 & 0 & 0 & 0 \\
\end{array}
\right)$ & & $\left(
\begin{array}{ccccc}
 0 & 0 & 0 & 0 & 0 \\
 0 & 0 & \frac{i \kappa}{2 f} & 0 & 0 \\
 0 & -\frac{i \kappa}{2 f} & 0 & 0 & 0 \\
 0 & 0 & 0 & 0 & 0 \\
 0 & 0 & 0 & 0 & 0 \\
\end{array}
\right)$ \\

$\chi^0_5$ & & $\left(
\begin{array}{ccccc}
 0 & \frac{i y_L c_\theta }{\sqrt{3}} & \frac{i y_L c_\theta}{\sqrt{3}} & 0 & 0 \\
 \frac{i y_R s_\theta}{\sqrt{3}} & 0 & 0 & 0 & 0 \\
 \frac{i y_R s_\theta}{\sqrt{3}} & 0 & 0 & 0 & 0 \\
 0 & 0 & 0 & 0 & 0 \\
 0 & 0 & 0 & 0 & 0 \\
\end{array}
\right)$ & & $\left(
\begin{array}{ccccc}
 0 & 0 & 0 & 0 & 0 \\
 0 & -\frac{\kappa}{2 \sqrt{3} f} & -\frac{\kappa}{2 \sqrt{3} f} & 0 & 0 \\
 0 & -\frac{\kappa}{2 \sqrt{3} f} & -\frac{\kappa}{2 \sqrt{3} f} & 0 & 0 \\
 0 & 0 & 0 & 0 & \frac{\kappa}{\sqrt{3} f} \\
 0 & 0 & 0 & \frac{\kappa}{\sqrt{3} f} & 0 \\
\end{array}
\right)$ \\

$\chi^0_1$ & & $\left(
\begin{array}{ccccc}
 0 & \frac{i y_L (c_\theta-3)}{2 \sqrt{6}} & \frac{i y_L (c_\theta+3)}{2 \sqrt{6}} & 0 & 0 \\
 \frac{i y_R s_\theta}{2 \sqrt{6}} & 0 & 0 & 0 & 0 \\
 \frac{i y_R s_\theta}{2 \sqrt{6}} & 0 & 0 & 0 & 0 \\
 0 & 0 & 0 & 0 & 0 \\
 0 & 0 & 0 & 0 & 0 \\
\end{array}
\right)$ & & $\left(
\begin{array}{ccccc}
 0 & 0 & 0 & 0 & 0 \\
 0 & \frac{\kappa}{2 \sqrt{6} f} & -\frac{\kappa}{\sqrt{6} f} & 0 & 0 \\
 0 & -\frac{\kappa}{\sqrt{6} f} & \frac{\kappa}{2 \sqrt{6} f} & 0 & 0 \\
 0 & 0 & 0 & 0 & -\frac{\kappa}{\sqrt{6} f} \\
 0 & 0 & 0 & -\frac{\kappa}{\sqrt{6} f} & 0 \\
\end{array}
\right)$ \\

$\eta$ & & $\left(
\begin{array}{ccccc}
 0 & i \sqrt{\frac{5}{2}} y_L c^2_{\theta/2} & -i
   \sqrt{\frac{5}{2}} y_L s^2_{\theta/2} & 0 & 0 \\
 \frac{iy_R}{2} \sqrt{\frac{5}{2}}  s_\theta & 0 & 0 & 0 & 0 \\
 \frac{iy_R}{2} \sqrt{\frac{5}{2}} s_\theta & 0 & 0 & 0 & 0 \\
 0 & 0 & 0 & 0 & 0 \\
 0 & 0 & 0 & 0 & 0 \\
\end{array}
\right)$ & & $\left(
\begin{array}{ccccc}
 0 & 0 & 0 & 0 & 0 \\
 0 & \frac{3 \kappa}{2 \sqrt{10} f} & 0 & 0 & 0 \\
 0 & 0 & \frac{3 \kappa}{2 \sqrt{10} f} & 0 & 0 \\
 0 & 0 & 0 & -\frac{\kappa}{\sqrt{10} f} & 0 \\
 0 & 0 & 0 & 0 & -\frac{\kappa}{\sqrt{10} f} \\
\end{array}
\right)$ \\
\hline\hline
    \end{tabular}
\end{footnotesize}
\end{center}
    \caption{\sf\it The expressions for the couplings to the neutral pNGBs, $\mathcal{I}_{\pi^0_i}$ and  $\mathcal{I}_{\partial\pi^0_i}$ where $\pi_i^0\equiv h,\chi^0_3,\chi^0_5,\chi^0_1,\eta$.}
    \label{tab:neutral_int}
\end{table}

\begin{table}[t!]
\begin{center}
\begin{footnotesize}
    \centering
    \begin{tabular}{lllllllll}
    \hline\hline
        $\pi^+_j$ & & \thead{$\mathcal{I}^1_{\pi^+_j}$} & & \thead{$\mathcal{I}^2_{\pi^+_j}$} & & \thead{$\mathcal{I}_{\partial\pi^+_j}$} \\
        \hline
        $\chi^+_3$ & & $\left(
\begin{array}{cccc}
 0 & \frac{i y_L}{\sqrt{2}} & 0 & 0 \\
 0 & 0 & 0 & 0 \\
 0 & 0 & 0 & 0 \\
 0 & 0 & 0 & 0 \\
 0 & 0 & 0 & 0 \\
\end{array}
\right)$ & & $\left(
\begin{array}{cccc}
 0 & 0 & 0 & 0 \\
 -\frac{i y_L c_\theta}{\sqrt{2}} & 0 & 0 & 0 \\
 \frac{i y_L c_\theta}{\sqrt{2}} & 0 & 0 & 0 \\
 0 & 0 & 0 & 0 \\
 0 & 0 & 0 & 0 \\
\end{array}
\right)$ & & $\left(
\begin{array}{cccc}
 0 & 0 & 0 & 0 \\
 0 & -\frac{\kappa}{2 \sqrt{2} f} & 0 & 0 \\
 0 & \frac{\kappa}{2 \sqrt{2} f} & 0 & 0 \\
 0 & 0 & -\frac{i \kappa}{2 f} & 0 \\
 0 & 0 & 0 & \frac{i \kappa}{2 f} \\
\end{array}
\right)$ \\
        
        $\chi^+_5$ & & $\left(
\begin{array}{cccc}
 0 & \frac{y_L c_\theta}{\sqrt{2}} & 0 & 0 \\
 0 & 0 & 0 & 0 \\
 0 & 0 & 0 & 0 \\
 0 & 0 & 0 & 0 \\
 0 & 0 & 0 & 0 \\
\end{array}
\right)$ & & $\left(
\begin{array}{cccc}
 0 & -\frac{y_R s_\theta}{\sqrt{2}} & 0 & 0 \\
 -\frac{y_L c_\theta}{\sqrt{2}} & 0 & 0 & 0 \\
 -\frac{y_L c_\theta}{\sqrt{2}} & 0 & 0 & 0 \\
 0 & 0 & 0 & 0 \\
 0 & 0 & 0 & 0 \\
\end{array}
\right)$ & & $\left(
\begin{array}{cccc}
 0 & 0 & 0 & 0 \\
 0 & \frac{i \kappa}{2 \sqrt{2} f} & 0 & 0 \\
 0 & \frac{i \kappa}{2 \sqrt{2} f} & 0 & 0 \\
 0 & 0 & -\frac{\kappa}{2 f} & 0 \\
 0 & 0 & 0 & -\frac{\kappa}{2 f} \\
\end{array}
\right)$ \\
\hline\hline
        $\pi^+_j$ & & \thead{$\hat{\mathcal{I}}^1_{\pi^+_j}$} & & \thead{$\hat{\mathcal{I}}^2_{\pi^+_j}$} & & \thead{$\hat{\mathcal{I}}_{\partial\pi^+_j}$} \\
        \hline
        $\chi^+_3$ & & $\left(
\begin{array}{ccccc}
 -\frac{i y_L}{\sqrt{2}} & 0 & 0 & 0 & 0 \\
 0 & 0 & 0 & 0 & 0 \\
 0 & 0 & 0 & 0 & 0 \\
\end{array}
\right)$ & & $\left(
\begin{array}{ccccc}
 0 & 0 & 0 & 0 & 0 \\
 0 & 0 & 0 & 0 & 0 \\
 0 & 0 & 0 & 0 & 0 \\
\end{array}
\right)$ & & $\left(
\begin{array}{ccccc}
 0 & -\frac{\kappa}{2 \sqrt{2} f} & \frac{\kappa}{2 \sqrt{2} f} & 0 & 0 \\
 0 & 0 & 0 & \frac{i\kappa}{2 f} & 0 \\
 0 & 0 & 0 & 0 & -\frac{i\kappa}{2 f} \\
\end{array}
\right)$ \\
        
        $\chi^+_5$ & & $\left(
\begin{array}{ccccc}
 \frac{y_L c_\theta}{\sqrt{2}} & 0 & 0 & 0 & 0 \\
 0 & 0 & 0 & 0 & 0 \\
 0 & 0 & 0 & 0 & 0 \\
\end{array}
\right)$ & & $\left(
\begin{array}{ccccc}
 -\frac{y_R s_\theta}{\sqrt{2}} & 0 & 0 & 0 & 0 \\
 0 & 0 & 0 & 0 & 0 \\
 0 & 0 & 0 & 0 & 0 \\
\end{array}
\right)$ & & $\left(
\begin{array}{ccccc}
 0 & -\frac{i \kappa}{2 \sqrt{2} f} & -\frac{i \kappa}{2 \sqrt{2} f} & 0 & 0 \\
 0 & 0 & 0 & \frac{\kappa}{2 f} & 0 \\
 0 & 0 & 0 & 0 & \frac{\kappa}{2 f} \\
\end{array}
\right)$ \\
\hline\hline         
    \end{tabular}
\end{footnotesize}
\end{center}
    \caption{\sf\it The expressions for the couplings to the charged pNGBs, $\mathcal{I}^{1,2}_{\pi^+_j}$, $\hat{\mathcal{I}}^{1,2}_{\pi^+_j}$, $\mathcal{I}_{\partial\pi^+_j}$ and $\hat{\mathcal{I}}_{\partial\pi^+_j}$  where  $\pi^+_j\equiv \chi^+_3,\chi^+_5$.}
    \label{tab:charged_int}
\end{table}
\begin{table}[t!]
\begin{center}
\begin{footnotesize}
    \centering
    \begin{tabular}{lllllllll}
    \hline\hline
        & \thead{$\mathcal{I}^L_{Z}$} & \thead{$\mathcal{I}^R_{Z}$} \\
        \hline
        $Z$ \!\!\!\!\! & $\left(\!\!\!
\begin{array}{ccccc}
 \frac{3 c_W^2-s_W^2}{6} \!\!\!& 0 \!\!\!& 0 \!\!\!& 0 \!\!\!& 0 \\
 0 \!\!\!& \frac{3 c_{\theta }-4 s_W^2}{6} \!\!\!& 0 \!\!\!& -\frac{\kappa  s_{\theta }}{8}
   \!\!\!& \frac{\kappa  s_{\theta }}{8} \\
 0 \!\!\!& 0 \!\!\!& -\frac{4 s_W^2+3 c_{\theta }}{6} \!\!\!& \frac{\kappa  s_{\theta }}{8}
   \!\!\!& -\frac{\kappa  s_{\theta }}{8} \\
 0 \!\!\!& -\frac{\kappa  s_{\theta }}{8} \!\!\!& \frac{\kappa  s_{\theta }}{8} \!\!\!& -\frac{2 s_W^2}{3}
   \!\!\!& 0 \\
 0 \!\!\!& \frac{\kappa  s_{\theta }}{8} \!\!\!& -\frac{\kappa  s_{\theta }}{8} \!\!\!& 0 \!\!\!& -\frac{2
   s_W^2}{3} \\
\end{array}
\!\!\!\right)$  \!& $\left(\!\!\!
\begin{array}{ccccc}
 -\frac{2 s_W^2}{3} \!\!\!& 0 \!\!\!& 0 \!\!\!& 0 \!\!\!& 0 \\
 0 \!\!\!& \frac{3 c_{\theta }-4 s_W^2}{6} \!\!\!& 0 \!\!\!& -\frac{\kappa  s_{\theta }}{8}
   \!\!\!& \frac{\kappa  s_{\theta }}{8} \\
 0 \!\!\!& 0 \!\!\!& -\frac{4 s_W^2+3 c_{\theta }}{6} \!\!\!& \frac{\kappa  s_{\theta }}{8}
   \!\!\!& -\frac{\kappa  s_{\theta }}{8} \\
 0 \!\!\!& -\frac{\kappa  s_{\theta }}{8} \!\!\!& \frac{\kappa  s_{\theta }}{8} \!\!\!& -\frac{2 s_W^2}{3}
   \!\!\!& 0 \\
 0 \!\!\!& \frac{\kappa  s_{\theta }}{8} \!\!\!& -\frac{\kappa  s_{\theta }}{8} \!\!\!& 0 \!\!\!& -\frac{2
   s_W^2}{3} \\
\end{array}
\!\!\!\right)$\\
        \hline\hline 
      & \thead{$\mathcal{I}^L_{W}$} & \thead{$\mathcal{I}^R_{W}$} \\
        \hline 
$W^+$ \!\!\!\!\!& $\left(\!\!\!
\begin{array}{cccc}
 \frac{1}{\sqrt{2}} & 0 & 0 & 0 \\
 0 & \frac{c_{\theta }+1}{2 \sqrt{2}} & 0 & \frac{i \kappa  s_{\theta }}{4} \\
 0 & \frac{s^2_{\theta/2}}{\sqrt{2}} & \frac{i \kappa 
   s_{\theta }}{4} & 0 \\
 0 & -\frac{\kappa  s_{\theta }}{4 \sqrt{2}} & 0 & \frac{ i \left(c_{\theta
   }-1\right)}{2} \\
 0 & -\frac{\kappa  s_{\theta }}{4 \sqrt{2}} & -\frac{ i \left(c_{\theta }+1\right)}{2}
   & 0 \\
\end{array}
\!\!\!\right)$ \!& $\left(\!\!\!\begin{array}{cccc}
 0 & 0 & 0 & 0 \\
 0 & \frac{c_{\theta }+1}{2 \sqrt{2}} & 0 & \frac{i \kappa  s_{\theta }}{4} \\
 0 & \frac{s^2_{\theta/2}}{\sqrt{2}} & \frac{i \kappa 
   s_{\theta }}{4} & 0 \\
 0 & -\frac{\kappa  s_{\theta }}{4 \sqrt{2}} & 0 & \frac{i \left(c_{\theta
   }-1\right)}{2} \\
 0 & -\frac{\kappa  s_{\theta }}{4 \sqrt{2}} & -\frac{ i \left(c_{\theta }+1\right)}{2}
   & 0 \\
\end{array}
\!\!\!\right)$ \\
\hline\hline 
      & \thead{$\hat{\mathcal{I}}^L_{W}$} & \thead{$\hat{\mathcal{I}}^R_{W}$} \\
        \hline 
$W^+$ \!\!\!\!\!& $\left(\!\!\!
\begin{array}{ccccc}
 0 & \frac{s^2_{\theta/2}}{\sqrt{2}} & \frac{c_{\theta }+1}{2
   \sqrt{2}} & -\frac{\kappa  s_{\theta }}{4 \sqrt{2}} & -\frac{\kappa  s_{\theta }}{4
   \sqrt{2}} \\
 0 & \frac{i \kappa  s_{\theta }}{4}  & 0 & 0 & -\frac{i \left(c_{\theta
   }+1\right)}{2} \\
 0 & 0 & \frac{i \kappa  s_{\theta }}{4} & \frac{i \left(c_{\theta }-1\right)}{2} &
   0 \\
\end{array}
\!\!\!\right)$ \!& $\left(\!\!\!
\begin{array}{ccccc}
 0 & \frac{s^2_{\theta/2}}{\sqrt{2}} & \frac{c_{\theta }+1}{2
   \sqrt{2}} & -\frac{\kappa  s_{\theta }}{4 \sqrt{2}} & -\frac{\kappa  s_{\theta }}{4
   \sqrt{2}} \\
 0 & \frac{i \kappa  s_{\theta }}{4} & 0 & 0 & -\frac{i \left(c_{\theta
   }+1\right)}{2} \\
 0 & 0 & \frac{i \kappa  s_{\theta }}{4} & \frac{i \left(c_{\theta }-1\right)}{2} &
   0 \\
\end{array}
\!\!\!\right)$  \\
\hline\hline
    \end{tabular}
\end{footnotesize}
\end{center}
    \caption{\sf\it The expressions for the couplings to the $Z$ and $W$ bosons, $\mathcal{I}^{L,R}_{Z}$, $\mathcal{I}^{L,R}_{W}$, and $\hat{\mathcal{I}}^{L,R}_{W}$.}
    \label{tab:gauge_int}
\end{table}
Note that both the Lagrangians above are obtained from Eqs.\,\eqref{L_elem}, \eqref{L_tp} and \eqref{L_PC} together with the choice given in Eq.\,\eqref{discretechoice}. We only keep terms with one pNGB (or gauge boson) and at least one fermion with $Q=2/3$ in Eqs.\,\eqref{L_int} and \eqref{L_int_gauge}. 
The mass matrix $\mathcal{M}_{2/3}$ is given by
\begin{align}
\mathcal{M}_{2/3}=\left(
\begin{array}{ccccc}
 0 & 0 & 0 & -\frac{y_L f}{2} s_\theta & \frac{y_L f}{2} s_\theta \\
 0 & M & 0 & 0 & 0 \\
 0 & 0 & M & 0 & 0 \\
 -\frac{y_R f}{2}  (c_\theta +1) & 0 & 0 & M & 0 \\
 -y_R f s^2_{\theta/2} & 0 & 0 & 0 & M \\
\end{array}
\right),  
\end{align}
while the interaction matrices $\mathcal{I}_{\pi^0_i}$, $\mathcal{I}_{\partial\pi^0_i}$, $\mathcal{I}^{1,2}_{\pi^+_j}$, $\hat{\mathcal{I}}^{1,2}_{\pi^+_j}$, $\mathcal{I}_{\partial\pi^+_j}$ and $\hat{\mathcal{I}}_{\partial\pi^+_j}$ are given in the Tables\,\ref{tab:neutral_int} and \ref{tab:charged_int}, respectively. 
The interaction matrices involving the the weak gauge bosons $\mathcal{I}^{L,R}_Z$, $\mathcal{I}^{L,R}_W$ and $\hat{\mathcal{I}}^{L,R}_W$ are given in Table\,\ref{tab:gauge_int}. The fermion mass basis are obtained through singular value decomposition of $\mathcal{M}_{2/3}$ using biunitary rotations $\mathcal{T}_{L,R}\to U_{L,R}\mathcal{T}_{L,R}$. The expressions for the matrices $U_{L,R}$ at the leading order in $\theta$ are given by
\begin{align}
U_L=\left(
\begin{array}{ccccc}
 0 & 0 & 1 & 0 & 0 \\
 0 & 1 & 0 & 0 & 0 \\
 -1 & 0 & 0 & -\frac{f M y_L \theta }{2 \left(M^2+f^2 y_R^2\right)} &
   \frac{f y_L \theta }{2 M} \\
 \frac{f y_L \theta }{2 M} & 0 & 0 & 0 & 1 \\
 \frac{f M y_L \theta }{2 M^2+2 f^2 y_R^2} & 0 & 0 & -1 & 0 \\
\end{array}
\right),\,
U_R=\left(
\begin{array}{ccccc}
 0 & 0 & 1 & 0 & 0 \\
 0 & 1 & 0 & 0 & 0 \\
 \frac{M}{\sqrt{M^2+f^2 y_R^2}} & 0 & 0 & \frac{f y_R}{\sqrt{M^2+f^2
   y_R^2}} & 0 \\
 0 & 0 & 0 & 0 & 1 \\
 \frac{f y_R}{\sqrt{M^2+f^2 y_R^2}} & 0 & 0 & -\frac{M}{\sqrt{M^2+f^2
   y_R^2}} & 0 \\
\end{array}
\right).
\end{align}
Similarly for $\mathcal{B}$ the mass matrix $\mathcal{M}_{-1/3}$ is given by
\begin{align}
\mathcal{M}_{-1/3}=\left(
\begin{array}{cccc}
 \mu _b\theta & 0 & -\frac{i f y_L s_\theta}{\sqrt{2}} & \frac{i f y_L s_\theta}{\sqrt{2}} \\
 0 & M & 0 & 0 \\
 0 & 0 & M & 0 \\
 0 & 0 & 0 & M \\
\end{array}
\right).
\end{align}
while the biunitary rotation matrices, defined by $\mathcal{B}_{L,R}\to \tilde{U}_{L,R}\mathcal{B}_{L,R}$\,, are given at the $\mathcal{O}(\theta)$ as
\begin{align}
\tilde{U}_{L}=\left(
\begin{array}{cccc}
 1 & 0 & \frac{if y_L \theta }{\sqrt{2} M} & -\frac{if y_L \theta }{\sqrt{2}
   M} \\
 0 & 0 & \frac{1}{\sqrt{2}} & \frac{1}{\sqrt{2}} \\
 0 & 1 & 0 & 0 \\
 -\frac{i f y_L\theta }{M} & 0 & -\frac{1}{\sqrt{2}} & \frac{1}{\sqrt{2}} \\
\end{array}
\right),\, \quad 
\tilde{U}_{R}=\left(
\begin{array}{cccc}
 1 & 0 & 0 & 0 \\
 0 & 0 & \frac{1}{\sqrt{2}} & \frac{1}{\sqrt{2}} \\
 0 & 1 & 0 & 0 \\
 0 & 0 & -\frac{1}{\sqrt{2}} & \frac{1}{\sqrt{2}} \\
\end{array}
\right).    
\end{align}

\section{Branching ratios involving degenerate states}
\label{decaywidths}

As mentioned in the section\,\ref{pheno}, in the presence of $N_\mathcal{T}$ nearly degenerate top-partners the cross section for the process $pp\to {\mathcal{T}}\overline{{\mathcal{T}}}\to A \bar B$ shown in Fig.\,\ref{fig:feymandiag} can be factorized as 
\begin{align}
\sigma(pp\to\mathcal{T}\overline{{\mathcal{T}}}\to A\bar B)\overset{\mathrm{NWA}}{=}N_\mathcal{T}\sigma(pp\to \mathcal{T}\overline{{\mathcal{T}}}) \mathcal{BR}_2(\mathcal{T}\overline{{\mathcal{T}}}\to A \bar B)\, ,
\label{eq_BR2_app}
\end{align}
where $\sigma(pp\to \mathcal{T}\overline{{\mathcal{T}}})$ is the pair production cross section, which is the same for each partner, and $\mathcal{BR}_2({\mathcal{T}}\overline{{\mathcal{T}}}\to A \bar B)$ denotes the joint branching ratio of $\mathcal{T}\to A$ and $\overline{\mathcal{T}}\to \bar{B}$. 
The full cross section \eqref{eq_BR2_app} arises from squaring the amplitude in Fig.\,\ref{fig:feymandiag} and contains a term
\begin{equation}
  {\mathrm{tr}}\left[\Delta(p_1^2)^\dagger \Gamma^A(p_1^2) \Delta(p_1^2) \Delta(p_2^2) \Gamma^{\bar B}(p_2^2) \Delta(p_2^2)^\dagger\right], \label{BR2def}
\end{equation}
where the trace is over the $N_\mathcal{T}$ states and $p_{1,2}$ are the momenta of the two fermionic partners to be integrated over. The $\Delta$s are defined in Eq.\,\eqref{matrix_prop} with a total matrix-valued width $\Gamma_\mathcal{T}=\sum_A \Gamma^A(M^2_\mathcal{T})\equiv \sum_{\bar B} \Gamma^{\bar B}(M^2_\mathcal{T})$, where $\Gamma^A(M^2_\mathcal{T})$ and $\Gamma^{\bar B}(M^2_\mathcal{T})$ are the on-shell matrix-valued partial decay widths.

If it wasn't for the matrix-valued nature of the propagators and widths, one could further simplify Eq.\,\eqref{BR2def} by commuting the $\Delta$s and use the NWA to write the expression as the product of two ordinary branching ratios after integrating over $p_1^2$ and $p_2^2$. 
Here instead we need to treat each of the two pieces under the trace, namely $\Delta(p_1^2)^\dagger \Gamma^A(p_1^2) \Delta(p_1^2)$ and $\Delta(p_2^2) \Gamma^{\bar B}(p_2^2) \Delta(p_2^2)^\dagger$ as matrix-valued distributions in $p_1^2$ and $p_2^2$ respectively.

Let us consider the $p_1^2$ term for definiteness and define $\mathcal{Z} = \Gamma_\mathcal{T}+2i\delta M$. One can show that in the limit $\mathcal{Z}\sim 0$ where all components of $\mathcal{Z}$ become small compared to $M_\mathcal{T}$
\begin{align}
\label{int_1}
\int dp^2_1\, \Delta(p_1^2)^\dagger \Gamma^A(p_1^2) \Delta(p_1^2)\overset{\mathcal{Z}\sim 0}{\to}\frac{\pi}{M_\mathcal{T}} \sum_{n=1}^{2}\left(c_n^+\mathcal{O}_n^{A+}-ic_n^-\mathcal{O}_n^{A-}\right), 
\end{align}
where the matrix operators $\mathcal{O}^{A\pm}_n$ are given by
\begin{equation}
\mathcal{O}^{A\pm}_n\equiv \frac{1}{2}\left(\mathcal{Z}^{\dagger -n}\Gamma^A(M^2_\mathcal{T})\mathcal{Z}^{n-1}\pm \mathcal{Z}^{\dagger n-1}\Gamma^A(M^2_\mathcal{T})\mathcal{Z}^{-n}\right),
\end{equation}
and the coefficients $c^\pm_n$ only depend on $\mathcal{Z}$ and can be extracted by tracing the expression \eqref{int_1} over four linearly independent matrices, e.g. $\mathbb{1}_2,\sigma^a$.\footnote{One can find closed expressions for these coefficients in terms of traces of powers of $\mathcal{Z}$ and $\mathcal{Z}^\dagger$, but the expressions are rather unwieldy unless $\mathcal{Z}$ is hermitian. We will present further details in a forthcoming publication.}

Applying the above result to our problem we can now write the joint branching ratio as 
\begin{equation}
\mathcal{BR}_2({\mathcal{T}}\overline{{\mathcal{T}}}\to A \bar B)=\frac{1}{N_\mathcal{T}}{\rm tr}\left(\left[ \sum_{n=1}^{2}\left(c_n^+\mathcal{O}_n^{A+}-ic_n^-\mathcal{O}_n^{A-}\right)\right] .
\left[\sum_{m=1}^{2}\left(c_m^+\bar{\mathcal{O}}_m^{\bar{B}+}+ic_m^-\bar{\mathcal{O}}_m^{\bar{B}-}\right)\right]\right),    \label{BRfinal}
\end{equation}
where the matrix operator  $\bar{\mathcal{O}}^{\bar{B}\pm}_m$ is defined similarly as
\begin{equation}
\bar{\mathcal{O}}^{\bar{B}\pm}_m\equiv \frac{1}{2}\left(\mathcal{Z}^{-m}\Gamma^{\bar{B}}(M^2_\mathcal{T})\mathcal{Z}^{\dagger m-1}\pm \mathcal{Z}^{m-1}\Gamma^{\bar{B}}(M^2_\mathcal{T})\mathcal{Z}^{\dagger -m}\right)\,. 
\end{equation}
Note that in \eqref{BRfinal} we extract a factor $1/N_\mathcal{T}$ to normalize $\mathcal{BR}_2$ to one.
Solving for $c_n^\pm$ for our benchmark parameters in section\,\ref{pheno}, we find $c^+_1=1.30$, $c^+_2=-0.296$, $c^-_1=0.192$, and $c^-_2=-0.072$.

\bibliography{toppartner}
\bibliographystyle{JHEP}
\end{document}